# The CO₂ Profile and Analytical Model for the Pioneer Venus Large Probe Neutral Mass Spectrometer

**Rakesh Mogul[1,2], Sanjay S. Limaye[3], M. J. Way[4,5,6]**

[1] Chemistry & Biochemistry Department, California State Polytechnic University, Pomona, CA, USA
[2] Blue Marble Institute of Science, Seattle, WA USA
[3] Space Science and Engineering Center, University of Wisconsin, Madison, WI, USA
[4] NASA Goddard Institute for Space Studies, 2880 Broadway, New York, NY, USA
[5] GSFC Sellers Exoplanet Environments Collaboration, Greenbelt, MD, USA
[6] Theoretical Astrophysics, Department of Physics and Astronomy, Uppsala University, Uppsala, Sweden

**\*Corresponding author:** Rakesh Mogul (rmogul@cpp.edu)

**Highlights**
- We re-analyzed archived mass spectra from the Pioneer Venus Large Probe and obtain novel chemical information regarding Venus' atmosphere.
- Our analytical model accounts for the impacts of descent and instrument configuration and enables the disentanglement of isobaric species.
- We obtained the first and most complete $CO_2$ altitude profile in units of density (kg m$^{-3}$) across ~55-1 km.
- Our volumetric abundances across 55.4-0.9 km of 96.6 ± 2.4% w/w $CO_2$ are consistent with associated measures from the Pioneer Venus Large Probe.
- Tracking of the $CO_2$ profile reveals hitherto unreported partial and rapidly clearing clogs of the LNMS at ≤17 km.
- The partial clogs are suggestive of a deep lower atmospheric haze at ≤17 km.

**Abstract**

We present a significantly updated $CO_2$ altitude profile for Venus (64.2-0.9 km) and provide support for a potential deep lower atmospheric haze of particles (≤17 km). We extracted this information by developing a new analytical model for mass spectra obtained by the Pioneer Venus Large Probe (PVLP) Neutral Mass Spectrometer (LNMS). Our model accounts for changes in LNMS configuration and output during descent and enables the disentanglement of isobaric species via a data fitting routine that adjusts for mass-dependent changes in peak shape. The model yields $CO_2$ in units of density (kg m$^{-3}$), isotope ratios for $^{13}C/^{12}C$ and $^{18}O/^{16}O$, and 14 measures of $CO_2$ density across 55.4-0.9 km, which represents the most complete altitude profile for $CO_2$ at ≤60 km to date. The $CO_2$ density profile is also consistent with the pressure, temperature, and volumetric gas measurements from the PVLP and VeNeRa spacecraft. Nominal and low-noise operations for the LNMS mass analyzer are supported by the behaviors (*e.g.*, ionization yields, fragmentation yields, and peak shapes) of several internal standards (*e.g.*, $CH_3^+$, $CH_4^+$, $^{40}Ar^+$, $^{136}Xe^{2+}$, and $^{136}Xe^+$), which were tracked across the descent. Lastly, our review of the $CO_2$ profile and LNMS spectra reveals hitherto unreported partial and rapidly clearing clogs of the inlet in the lower atmosphere, along with several ensuing data spikes at multiple masses. Together, these observations suggest that atmospheric intake was impacted by particles at ≤17 km and that rapid particle degradation at the inlet yielded a temporary influx of mass signals into the LNMS.

**1. Introduction**

Venus is again emerging as an exciting target for planetary exploration with multiple upcoming missions including DAVINCI and VERITAS from NASA (Garvin et al., 2022, Smrekar et al., 2022), EnVision from ESA (Ghail, 2021), Shukrayaan-1 from the Indian Space Research Organization (ISRO) (Antonita et al., 2022, Sundararajan, 2021), and a potential orbiter mission from the China National Space Administration (CNSA).  Likewise, astrobiology considerations for Venus have a gained substantial interest with the 2023-2032 Decadal Report highlighting that "Venus's atmosphere has been suggested as a prospective abode for life" with a potential habitable zone in the middle clouds.

The clouds and lower atmosphere of Venus were initially studied by descent probes as part of the Pioneer Venus (1978), VeGa (1984), and VeNeRa (1961-1984) programs.  The clouds and lower atmosphere are also subjects of the upcoming DAVINCI mission, which is scheduled to launch in 2029.  Hence, in the projected ~45 years between launches to Venus, the roles of archival data analysis, remote observations, ground-based testing, and computational approaches remain as the drivers for Venus science today.

In this context, we present a new analysis and interpretation of the mass spectra obtained by the Pioneer Venus (PV) Large Probe Neutral Mass Spectrometer (LNMS), which descended through Venus' atmosphere as an instrument on board the Pioneer Venus Large Probe (PVLP) on December 9, 1978.  Results from the LNMS were first reported in Hoffman, Hodges, et al. (1979), while the mixing ratios for major and minor atmospheric gases (based on $^{36}$Ar), and the associated analytical strategies, were described in Hoffman, Hodges, et al. (1979) and Hoffman, Hodges, Donahue, et al. (1980).  Design of the LNMS and selected control experiments were briefly described in Hoffman, Hodges, et al. (1979) and Hoffman, Hodges, Wright, et al. (1980), while a majority of the associated engineering, pre-flight testing, and in-flight diagnostic information remain in restricted documents in the PV mission archive at the NASA Ames Research Center (ARC).

Several crucial interpretations regarding Venus' atmosphere were obtained through the original LNMS investigations. Examples include assessments of the aerosols in the middle and lower clouds (56.5-47.5 km) being composed of hydrated sulfuric acid ($H_2SO_4(H_2O)_n$) and possessing a deuterium to hydrogen ratio (D/H = 0.016 ± 0.002) which is ~100-fold higher than the Standard Mean Ocean Water (Donahue et al., 1982, Donahue and Hodges, 1992, Donahue and Hodges, 1993, Hoffman, Hodges, Donahue, et al., 1980). Serendipitously, the data supporting these specific interpretations were obtained when the LNMS inlets were temporarily clogged by atmospheric aerosols between the altitudes of ~50-25 km. Under the conditions of the descent (*e.g.*, increasing temperature), these trapped aerosols eventually vaporized and dissociated at the inlets to yield an influx, or spikes, of chemical species within the LNMS (*e.g.*, $SO_2^+$, $SO^+$, $H_2O^+$, and $HDO^+$). In the original LNMS investigations, these mass signals were interpreted as chemical signatures for hydrated sulfuric acid.

Additionally, in Mogul et al. (2021), we showed that the LNMS data possess several *unassigned* mass positions and *uncharacterized* altitude trends. For example, our re-analyses of data from the middle clouds (55.4-51.3 km) support the presence of reduced chemicals including hydrogen sulfide ($H_2S$), water ($H_2O$), nitrous acid ($HNO_2$), and potentially phosphine ($PH_3$), ammonia ($NH_3$), and chlorous acid ($HClO_2$), which together are supportive of redox disequilbria within the clouds.

In this report, we expand on our re-analysis of the LNMS data and describe the development of an analytical model that includes the total mass spectra across the descent profile of 64.2-0.2 km. Our model includes a data fitting routine that remains responsive to changes in LNMS output and configuration during descent through the clouds and lower atmosphere. Herein, we characterize the behavior of several internal standards in the dataset, compare the trends to ground-based studies, and demonstrate nominal and low-noise operations of the LNMS mass analyzer during descent. We also describe extraction of the altitude profile for carbon dioxide ($CO_2$) in units of density (kg m$^{-3}$) and reveal the presence of hitherto unreported partial and rapidly clearing clogs of the LNMS at ≤17 km, which together illuminate the potential presence of a deep lower atmospheric haze at ≤17 km.

## 2. Data and Methods

### *2.1 Overview of the LNMS Operations*

The LNMS was a magnetic sector mass analyzer that measured ion count rates at 232 pre-selected and non-uniformly distributed mass positions across 1-208 amu. The LNMS mass scale in this report is interpreted as 1-208 u (unified atomic mass unit), or 1-208 Da (Dalton), and expressed herein as *m/z* 1-208. Our review of the LNMS data indicates a $^{12}$C-based mass scale (Prohaska et al., 2022), as many of the pre-selected mass positions correspond to the exact masses for several gases (*e.g.*, $SO_2$, $CO$, $NH_3$, $N_2$, and $O_2$) and atoms (*e.g.*, S, O, and C).

Ion counts in the LNMS were detected using two simultaneously operating mass channels (*m/z* 1-16 and 15-208) with discrete dynode electron multipliers (Be-Cu, 16 stage, discreet dynode). The LNMS operation cycles, or mass sweeps (64 s/each) included stepped measurements (59 s) across the pre-selected mass positions and a re-calibration step (5 s) of the ion accelerator power supply, which reset the peak stepping routine (Donahue and Hodges, 1992, Hoffman, Hodges and Duerksen, 1979, Hoffman, Hodges, Donahue, et al., 1980). An onboard microprocessor (Intel 4004[1]) in the LNMS afforded control over the operations inclusive of the rapid peak-stepping, integration of the ion count rates over 235 ms, control of the electron energy and electron multiplier gain through use of quantitative commands, and in-flight corrections to the ion acceleration voltage through tracking of internal calibrants (Fimmel et al., 1995, Hoffman, Hodges, Wright, et al., 1980). For each mass sweep, the ion acceleration potential was maintained at a tolerance of ~0.02% using the internal calibrants, $CH_4$ and $^{136}$Xe, which were injected as a pre-mixed gas into the ion source during each re-calibration and tracked at *m/z* 136 ($^{136}Xe^+$), *m/z* 68 ($^{136}Xe^{2+}$), and *m/z* 15 ($CH_3^+$; produced from fragmentation of $CH_4^+$) (Donahue and Hodges, 1993, Hoffman, Hodges, Donahue, et al., 1980, Hoffman, Hodges, Wright, et al., 1980).

---

[1] https://www.intel.com/content/www/us/en/history/museum-story-of-intel-4004.html

During the initial stages of the descent (>65 km), the LNMS performed several mass sweeps while all inlets to the atmosphere were closed, inclusive of the in-flight corrections. This pre-sampling data includes three complete mass spectra (*m/z* 1-208) and one incomplete spectrum (*m/z* ≤40), which together measured the internal calibrants, associated ionized and fragmented products, and background chemical species. After completion of these pre-sampling sweeps, inflow of atmospheric gases into the LNMS was permitted by pyrotechnic activation of the breakseal cap, which opened two ceramic coated tantalum inlet tubes (3.2 mm diameter), which were crimped to restrict inflow rates (Fimmel et al., 1995, Hoffman, Hodges, Donahue, et al., 1980).

As indicated in control experiments in Hoffman, Hodges, Wright, et al. (1980), nominal intake rates through the inlets, due to differing wall widths, were ~0.1 nL s$^{-1}$ (primary inlet) and ~1 nL s$^{-1}$ (secondary inlet); where flow through the secondary inlet was controlled by a dedicated valve, labeled as Valve 1 in the associated block diagram (Fig. 4, Hoffman, Hodges, Wright, et al. (1980)). To maintain stable pressures or densities within the ion source, atmospheric inflow was regulated by closure of Valve 1 via a pre-programmed sequence at ~47 km (~1.6 bar) and through a Variable Conductance Valve (VCV) that gradually opened the ion source cavity to a chemical getter (zirconium-aluminum; G1 in the associated block diagram) as atmospheric pressure increased. Together, these procedures impacted intake rates (Hoffman, Hodges, Wright, et al., 1980), ion transmission or flow through the mass analyzer towards the detector (Hoffman, Hodges, Donahue, et al., 1980), and detector output (Hoffman, Hodges and Duerksen, 1979), consistent with our review of the data.

From an altitude of 64.2 km down to the surface, a total of 46 mass spectra were collected; inclusive of 39 complete spectra obtained at an electron energy of 70 eV, 1 incomplete spectrum at 70 eV from 0.2 km that was interrupted by surface impact, 3 spectra obtained at 30 eV, and another 3 spectra at 22 eV. Across the descent, sampling intervals between these measures decreased from 2.3 to 1.0 km between 64.2 and 49.4 km, increased to 3.2 and 3.0 km at 45.2 and 42.2 km, after parachute deployment at ~47 km, and then steadily decreased to 0.7 km near the surface.  For this study, we retain the common altitude scale from Hoffman, Hodges, Donahue, et al. (1980), Oyama et al. (1980), Fimmel et al. (1995), which is based on the data obtained by the Atmospheric Structure Investigation (Seiff et al., 1980) and differs slighlty from the accepted scale published in Seiff et al. (1985).

As part of a pre-programmed sequence, atmospheric gases were collected at ~62 km into the Isotope Ratio Measuring Cell (IRMC) for the purpose of obtaining noble gas abundances. After being scrubbed of active gases by molecular sieves and a chemical getter (zirconium-graphite), the IRMC sample was then injected into the ion source at ~48 km, where the mass positions for all noble gases (including the internal calibrants of $^{136}Xe^+$ and $^{136}Xe^{2+}$) subsequently show mild increases in counts at 45.2 km, with subsequent decreases towards 25.9 km.

During the descent the LNMS experienced a temporary clog at the inlets between ~50-25 km due to build-up from the aerosol particles in the middle and lower clouds (Hoffman, Hodges, Donahue, et al., 1980, Mogul et al., 2021).  As shown in **Fig. 1**, tracking of carbon dioxide ($CO_2^+$) readily reveals the clog to the inlets, where counts for $CO_2^+$ decrease by ~$10^4$-fold between ~50-45.2 km (~$10^2$ counts) – when compared to values at 51.3 km (~$10^6$ counts).  Then, counts for $CO_2^+$ begin to increase at 42.2 km and reach roughly expected values by 25.9 km.  Given the closure of Valve 1 (at ~47 km), the expected counts for $CO_2^+$ at ≤25 km are ≥$10^5$, as inferred from control experiments in Hoffman, Hodges and Duerksen (1979), which were presumably conducted with the pre-flight LNMS.

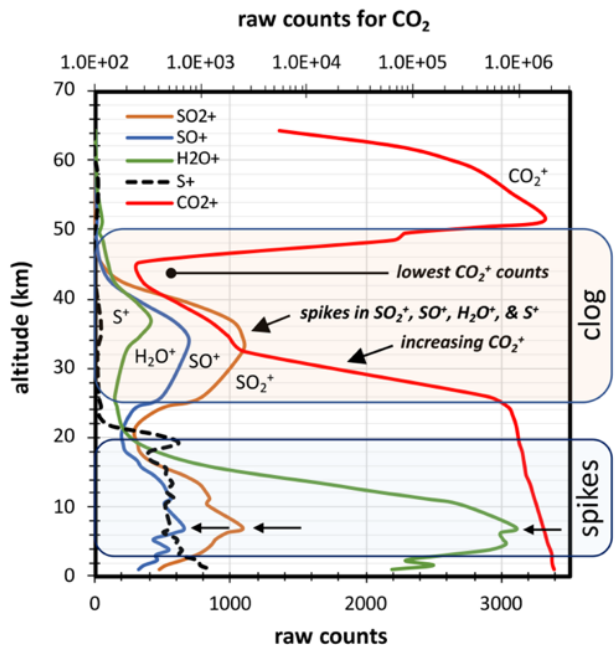

Fig. 1. Altitude profiles (64.2-0.9 km) for raw counts of $CO_2^+$ (*m/z* 44) which are described by the top x-axis, and $SO_2^+$ (*m/z* 64), $SO^+$ (*m/z* 48), $S^+$ (*m/z* 32), and $H_2O^+$ (*m/z* 18) which are described by the bottom x-axis; shaded rectangles (rounded edges) demark the altitudes for the main clog (~50-25 km) and deep lower atmosphere (≤17 km) where several mass positions exhibit spikes in the counts.

As the clogged inlet/s of the LNMS began to clear, several mass positions exhibited temporary increases or spikes in the counts. For example, as shown in **Fig. 1**, between 42.2 and 25.9 km, substantial though *temporary* increases in counts for $SO_2^+$ (*m/z* 64), $SO^+$ (*m/z* 48), $S^+$ (*m/z* 32), and $H_2O^+$ (*m/z* 18) are observed. In Hoffman, Hodges, Donahue, et al. (1980), the mass signals for $SO_2^+$, $SO^+$, and $H_2O^+$ were interpreted as signatures for hydrated sulfuric acid aerosols in the middle and lower clouds. In summary, during descent, the aerosols likely collected on the LNMS inlets and prevented intake of atmospheric gases – as evidenced by the large drop in $CO_2^+$ counts between ~50-25 km. As the descent continued, the hydrated sulfuric acid trapped at the inlets then vaporized and dissociated in the hotter lower atmosphere to yield an influx of chemical species in the LNMS (*e.g.*, $SO_2^+$, $SO^+$, $S^+$, and $H_2O^+$).

Between 24.4-20.3 km (**Fig. 1**), the LNMS data are consistent with a resumption of optimal atmospheric inflow. During this component of the descent, counts for $SO_2^+$, $SO^+$, and $H_2O^+$ remained at stable appreciable levels (>100 counts) and exhibited no spikes. Hence, these signals between 24.4-20.3 km are likely representative of atmospheric $SO_2$ and $H_2O$ (with $SO^+$ being a fragmentation product of $SO_2$), consistent with the interpretations in Hoffman, Hodges,

Donahue, et al. (1980).

Below ~17 km, however, our analyses show that $CO_2$ inflow is again impacted through several partial and rapidly clearly clogs. Evidence for these clogs include increases (or spikes) of ~5-500-fold in the counts across >37 mass signals between *m/z* 10-208, including *m/z* 64, 48, and 18. These LNMS data spikes were alluded to in Fimmel et al. (1995) as potentially being related to electrical and sensor anomalies that occurred at <12.5 km in each of the PV spacecraft. However, the dedicated investigations described in Seiff et al. (1995) found (or stated) no correlations between the LNMS operations and the PV electrical and sensor anomalies. In this absence of a tangible explanation, we show that the $CO_2$ density profile is supportive of the LNMS inlets being clogged at ≤17 km by a deep atmospheric haze of particles that ultimately degraded at the inlet to yield an influx of chemical species in the LNMS mass analyzer.

### *2.2 LNMS Data Analysis and Interpretation*

To construct an analytical model that accounts for LMNS configuration and output, the total spectra were divided into subsets and separately analyzed and interpreted. For example, mass spectra (70 eV) from the pre-sampling data represent measures of the closed LNMS (total number of spectra, n = 3-4) and were treated as measures of *in situ* repeatability for the LNMS measures. Mass spectra (70 eV) from the descent profile (n = 38; 64.2-0.9 km) were further sub-divided into data from the clouds (64.2-51.3 km), main clog (~50-25 km), and lower atmosphere (24.4-0.2 km) to account for changes in valve configuration and the substantial though temporary obstruction of atmospheric intake. Spectra from the low eV (30 and 22 eV) profiles were collated and analyzed together (30 eV: n = 3 at 54, 30.1, and 9.1 km; 22 eV: n = 3 at 52.7, 28.1, and 8.2 km).

Due to the non-uniform distribution of the LNMS mass positions in the data, many chemical species were represented by clusters of mass points (≥ 3 mass positions), while several other chemical species were represented by only one mass point. For the mass clusters, we employed a data fitting routine adapted from Mogul et al. (2021) to disambiguate isobaric species at each altitude – or species with similar masses that exhibit overlapping mass spectral peaks. Selected mass points were fit with Gauss functions using a custom-built macro (Solver; MS Excel) that iteratively performed the regressions across the pre-sampling, descent data, or low eV profiles. The regression solutions yielded peak amplitudes (fitted counts), locations (experimental mass), and peak widths (or standard deviation of the location). For this study, peak widths were expressed as the full width half maximum (FWHM), per the conversion from Weisstein (2002). Example species represented by one mass position in the LMNS data include $SO_2^+$ (63.962 amu), $SO^+$ (47.966 amu), $CO_2^+$ (43.991 amu), $CO^{18}O^+$ (45.995 amu), $^{13}CO_2^+$ (44.991 amu), $C^{18}O^+$ (29.997 amu), $^{13}CO_2^{2+}$ (22.496 amu), $CO_2^{2+}$ (21.995 amu), $^{14}N^+$ (14.000 amu), and $C^+$ (12.000 amu).

### 2.3 Internal Standards for the LNMS

To characterize the performance of the LNMS during operation, we identified and tracked the behavior of several internal standards in the dataset (**Table 1**). Trends regarding FWHM and mass accuracy were obtained by fitting the chosen internal peak standards of $CH_3^+$, $CO^+$, $N_2^+$, $^{40}Ar^+$, and $^{136}Xe^+$ to Gauss functions and tracking the results across the pre-sampling, descent, and low eV profiles. Trends regarding single and multiple ionization of a parent species were obtained by tracking $^{136}Xe^+$, $^{136}Xe^{2+}$, $^{136}Xe^{3+}$, $^{40}Ar^+$, and $^{40}Ar^{2+}$ across the pre-sampling, descent, and low eV data. Trends regarding fragmentation yields were obtained by tracking $CH_4^+$, $^{13}CH_4^+$, $CH_3^+$, $CH_2^+$, and $C^+$ across the pre-sampling data.

The internal peak standards were fit as described to obtain FWHM (peak width) and experimental mass (peak location or center). Regressions were performed for $CH_3^+$ (an internal calibrant, per **Section 2.1**) using five mass points at *m/z* 15 (*LNMS mass positions*: 15.013, 15.018, 15.023, 15.028, and 15.033 amu). Regressions were performed for $^{136}Xe^+$ (an internal calibrant, per **Section 2.1**) using seven mass points at *m/z* 136 (*LNMS mass positions*: 135.453, 135.742, 135.824, 135.907, 135.990, 136.073, and 136.448 amu). Regressions for $^{40}Ar^+$ were obtained using five mass points at *m/z* 40 (*LNMS mass positions*: 39.950, 39.958, 39.965, 39.972, and 40.029 amu). Sources of argon in the LNMS included terrestrial contamination (as indicated by counts in the pre-sampling data) and the Venus atmosphere (as inferred by increasing counts during the descent).

For $CO^+$ (a fragment of $CO_2^+$ and trace Venus atmospheric gas) and isobaric $N_2^+$ (a terrestrial contaminant and minor Venus atmospheric gas), fits were obtained using six mass points at *m/z* 28 (*LNMS mass positions*: 27.988, 27.995, 28.000, 28.005, 28.012, and 28.032 amu). For fits at *m/z* 28, the regressions were constrained by (1) obtaining initial estimates of $CO^+$ using counts of $C^{18}O^+$ at the respective altitude and the averaged $^{18}O/^{16}O$ ratio from $CO_2^+$, (2) placing upper and lower limits to $CO^+$ counts using the propagated error (1 σ) from conversion using the $^{18}O/^{16}O$ ratio, and (3) accounting for isobars to $C^{18}O^+$ using expanded *lower* limits (3 σ) between 20-10 km to prevent overestimation of $CO^+$. Abundances for $N_2^+$ will be presented in a separate report.

*Table 1.* List of internal standards chosen for this study, and internal calibrants used in the LNMS.

| Internal Standards and Internal Calibrants | |
|---|---|
| peak standards *(all spectra)* | $CH_3^+$, $CO^+$, $N_2^+$, $^{40}Ar^+$, $^{136}Xe^+$ |
| ionization standards *(all spectra)* | $^{136}Xe^+$, $^{136}Xe^{2+}$, $^{136}Xe^{3+}$, $^{40}Ar^+$, $^{40}Ar^{2+}$ |
| fragmentation standards *(pre-sampling spectra)* | $CH_4^+$, $CH_3^+$, $CH_2^+$, $C^+$, $H^+$ |
| calibrants *(all spectra)* | $CH_3^+$, $^{136}Xe^{2+}$, $^{136}Xe^+$ |

**2.3 Targeted Data Fitting**

Regressions for selected mass clusters (2-6 mass positions per cluster) were fit with Gauss functions (**Section 2.1**) and solved across the pre-sampling, descent, and/or low eV data. Regressions were minimized using the sum of the squared deviations and constrained by (1) obtaining the expected FWHM for the target species and isobar/s using the linear relationship between FWHM and *m/z* of the internal peak standards at each altitude (expressed in this report as an averaged value of $R^2$ = 0.998 ± 0.002 across 64.2-0.9 km), (2) placing upper and lower limits (≤2σ) on the expected FWHM as calculated using the propagated error (σ) obtained from the standard errors of the slopes and intercepts from the linear regressions between FWHM and *m/z* of the internal peak standards at each altitude (LINEST function, MS Excel), (3) permitting maximum shifts in the mass scale equal to the maximum mass errors ($\Delta(m/z)$) exhibited by the internal peak standards of $m/z \leq 40$ at each altitude, and (4) placing limits on isobar abundances using isotope or ionization ratios, when possible.  Errors or uncertainties from the fits were estimated using the square root of the variance, which was calculated by summation of the square of the residuals at each altitude.

Each minimized fit accounted for ≥95% of the counts included in the regression, which was calculated by expressing the absolute deviations at the respective mass positions in the cluster as a percent of the total included counts.  In certain circumstances, to achieve ≥95% coverage, differing species at varying altitudes required relaxed limits for the FWHM (3-4 $\sigma$) and/or mass scale (2-4x $\Delta(m/z)$).  To ascertain the quality of the fits across the descent, pre-sampling, and low eV profiles, the fraction of counts covered by the total regressions for the mass cluster (expressed as percent) in the respective profile were calculated by respectively obtaining the summed absolute deviations and total counts across the combined spectra.  Details for exemplar fits are provided below:

- At *m/z* 16, the total minimized regressions for $O^+$ and $CH_4^+$ using three mass points (*LNMS mass scale*: 15.995, 16.031, and 17.002 amu) across the descent *and* pre-sampling profiles (42 spectra) accounted for 99.4% of the included counts;

- At *m/z* 18, regressions for $H_2O^+$, $^{36}Ar^{2+}$, $^{18}O^+$, and $NH_4^+$ (and/or $NH_2D^+$) using three mass points (17.985, 18.010, and 18.034 amu) across the descent *and* pre-sampling profiles (42 spectra) accounted for 98.5% of the included counts; regression constraints were inclusive of initial estimates for $^{36}Ar^{2+}$ obtained from $^{36}Ar^+$ counts and the averaged ratio for $^{40}Ar^{2+}/^{40}Ar^+$ from the pre-sampling data (22.2 ± 2.3%; **Section 3.1**), and initial estimates for $^{18}O^+$ obtained from counts for $^{16}O^+$ and the averaged $^{18}O/^{16}O$ ratio from $CO_2$ (2.195 x $10^{-3}$ ± 0.156 x $10^{-3}$; **Section 3.8**);

- At *m/z* 20, regressions for $^{40}Ar^{2+}$, $^{20}Ne^+$, $HF^+$, and $H_2^{18}O^+$ using four mass points (19.981, 19.992, 20.006, 20.015 amu) across the descent *and* pre-sampling profiles (42 spectra) accounted for 99.4% of the included counts, regression constraints were inclusive of initial estimates for $^{40}Ar^{2+}$ obtained from fitted counts for $^{40}Ar^+$ and the averaged ratio for $^{40}Ar^{2+}/^{40}Ar^+$ from the pre-sampling data, and initial estimates for $H_2^{18}O^+$ obtained from fitted counts for $H_2O^+$ and the averaged $^{18}O/^{16}O$ ratio from $CO_2$;

- At *m/z* 28, regressions for $CO^+$ and $N_2$ in the descent *and* pre-sampling profiles (42 spectra), described in **Section 2.2**, accounted for 98.2% of the included counts;

- At *m/z* 29, regressions for $^{13}CO^+$, $^{14}N^{15}N^+$, and $C_2H_5^+$ using three mass points at *m/z* 29 (28.997, 29.003, and 29.039 amu) across the descent *and* pre-sampling profiles (42 spectra) accounted for 98.6% of the included counts; regression constraints included (1) initial estimates of $^{14}N^{15}N^+$ obtained using fitted counts for $N_2^+$ and the $^{15}N/^{14}N$ ratio from Mogul et al. (2021), (2) initial estimates of $^{13}CO^+$ obtained using fitted $CO^+$ counts and the $^{13}C/^{12}C$ ratio from $CO_2^+$ (1.260 x $10^{-2}$ ± 0.072 x $10^{-2}$; as detailed in **Section 3.8**), and (3) *upper* limits of 3 σ for $^{13}CO^+$ abundances, as estimated using the propagated error (σ) after conversion using the $^{13}C/^{12}C$ ratio, and expanded *upper* limits of 6 σ at 64.2, 42.2, and 39.3 km;

- At *m/z* 32, regressions for $^{32}S^+$ and $O_2^+$ using two mass points at *m/z* 32 (31.972 and 31.990 amu) across the descent profile (38 spectra) accounted for 99.7% of the included counts.

***2.4 Normalization of Raw Counts***

Raw counts were normalized to $^{136}Xe^{2+}$ to account for increases and variances in transmission through the spectrometer during descent. Intake rates through the inlets for co-migrating species were treated as equivalent due to viscous flow per Donahue and Hodges (1992). Target species were normalized using **Eqn. 1**, which adjusts ion counts (*I*) for a target species (*i*) at a specific altitude (*a*) using counts for $^{136}Xe^{2+}$ at the specific altitude ($(I_{Xe})_a$) and selected reference (*R*) counts for $^{136}Xe^{2+}$ ($(I_{Xe})_a^R$). For data obtained within the clouds (64.2 to 51.3 km), the reference counts ($(I_{Xe})_a^R$) were assigned at 61.9 km. For the lower atmosphere (24.4 to 0.2 km), averaged reference counts ($(I_{Xe})_a^R$) were assigned to 24.4, 23, and 21.6 km. For data collected during the main clog (50.3-25.9 km), normalizations were coarsely achieved using the ratio of $^{134}Xe^{2+}/^{136}Xe^{2+}$ from 49.4 km since the IRMC sequence negated use of the $^{136}Xe^{n+}$. Respective errors in the counts at all altitudes were propagated throughout the calculations. Normalizations to $^{136}Xe^+$ and $^{136}Xe^{3+}$ were included for comparative purposes.

$$(1) \quad (I_i)_a^N = (I_i)_a * \left(\frac{(I_{Xe})_a^R}{(I_{Xe})_a}\right)$$

*2.5 Conversion to Carbon Dioxide Density*

Calibration curves were constructed to convert counts of $CO_2$ in the LNMS to units of density (kg m$^{-3}$). Calibration curves (or standard curves) were assembled using Venus atmospheric pressures and temperatures from Seiff et al. (1985), volumetric measures of $CO_2$ obtained by the PV Gas Chromatograph (PVGC) from Oyama et al. (1980), and $CO_2$ control experiments from Hoffman, Hodges and Duerksen (1979), which were presumably performed using the preflight LNMS. To account for the pre-programmed closure of Valve 1 (~47 km), data from the clouds and lower atmosphere were independently converted to density. As reference values for the conversions, atmospheric and $CO_2$ densities were first obtained at the altitudes of 51.6 km (0.698 bar) and 21.6 km (17.8 bar), which correspond to where the PVGC measured $CO_2$ abundances of 95.4% and 96.4% v/v, respectively (Oyama et al., 1980). In turn, the volumetric percent values at 51.6 km (0.698 bar) and 21.6 km (17.8 bar) were converted using **Eqn. 2** (*MW*, molecular weight) into mass percent values of 97.0 and 97.7% w/w $CO_2$, respectively.

$$(2) \quad (\% \, w/w)^{CO_2} = \frac{(\%v/v * MW)^{CO_2}}{(\%v/v * MW)^{CO_2} + (\%v/v * MW)^{N_2}}$$

To obtain atmospheric densities at 51.6 km (0.698 bar) and 21.6 km (17.8 bar), we extracted atmospheric data from Seiff et al. (1985), constructed a non-linear regression model ($R^2 = \geq 0.999$) correlating bulk atmospheric density to atmospheric pressure, and used the resultant regression equation to obtain atmospheric densities at 0.698 (1.36 kg m$^{-3}$) and 17.8 bar (16.8 kg m$^{-3}$). In turn, $CO_2$ densities at 51.6 km (0.698 bar) and 21.6 km (17.8 bar), were obtained via multiplication of the respective atmospheric densities by the mass percent values of 97.0 and 97.7% w/w $CO_2$.

The *calculated* densities for $CO_2$ at 51.6 km (1.31 kg m$^{-3}$) and 21.6 km (16.4 kg m$^{-3}$) were then translated to LNMS detector output (amps) using control data extracted from Figure 3 in Hoffman, Hodges and Duerksen (1979) – which tracked changes in outputted current (amps) in the LNMS across a pressure-temperature profile for $CO_2$ that simulated the PV Large Probe descent, inclusive of closure of the secondary inlet via Valve 1 at ~1.6 bar, or ~47 km in the descent profile. After digital extraction of the data (WebPlotDigitizer, https://automeris.io/WebPlotDigitizer), the pressure and temperature values were converted to $CO_2$ densities using the equation of state.

Power regression models ($R^2$ = ≥0.987) were then applied to correlate $CO_2$ density to current (amps) for control data (Hoffman, Hodges and Duerksen, 1979) from before and after the valve closure. The respective regression equations were then used to obtain the *expected* amps for the *calculated* $CO_2$ densities at 51.6 and 21.6 km. As described in Hoffman, Hodges and Duerksen (1979), current output by the detector was converted to digital counts by the multiplier pre- and post-amplifiers. Therefore, to convert amps to counts, we obtained an estimate of the ratio of counts per amp (counts/amp) at ~51.6 and 21.6 km through division of the normalized counts for $CO_2$ at 51.3 and 21.6 km, as measured by the LNMS at Venus, by the calculated *expected* amps at 51.6 and 21.6 km, respectively.

Calibration curves were then constructed by converting the data (amps) from Figure 3 in Hoffman, Hodges and Duerksen (1979) into units of counts. The data obtained *before* the valve closure (≤1.6 bar) were converted using the counts/amp ratio from ~51.6 km (~0.698 bar). The data obtained *after* the valve closure (>1.6 bar) were converted using the counts/amp from 21.6 km (17.8 bar). In turn, the calibration curves were fit to power regression models and the resulting regression equations ($R^2$ = ≥0.985) were used to convert normalized counts for $CO_2$ from the LNMS to density (kg m$^{-3}$). Data from the clouds (≤1.6 bar) were converted using the calibration curve for *before* the valve closure. Data from the lower atmosphere (<1.6 bar) were converted using the calibration curve for *after* the valve closure. Errors from the conversions were estimated using the standard error of the y-estimate, which were obtained by log-log regression analysis of the calibration curves (LINEST function, MS Excel).

## 3. Results

### 3.1 Background Chemical Inventory within the LNMS

The background chemical inventory within the LNMS was characterized using the pre-sampling data. Assignments were performed for *m/z* ≤136; however, several positions at *m/z* 125, 127, and ≥138 (with counts of ~1-5) remained unassigned. Summarized in **Table 2** are the background chemical species inclusive of (1) the described calibrant gases ($CH_4$ and $^{136}Xe$) and related ionized and fragmented species, (2) isotopes of Xe arising from the $^{136}Xe$ calibrant (Donahue et al., 1981), with the respective isotope ratios provided in parentheses, (3) several high and low abundance terrestrial contaminants (Hoffman, Oyama, et al., 1980), and (4) chemical species potentially formed in situ within the LNMS. Displayed in **Fig. 2** are counts for the background species from the pre-sampling sweeps (n = 3-4), which are organized as clustered columns (labeled as pre 1-4) to represent variances in the measures.

Counts for $^4He^+$, $CH_3^+$, $O^+$, $CH_4^+$, $H_2O^+$, $^{36}Ar^{2+}$, $^{38}Ar^{2+}$, $^{40}Ar^{2+}$, $^{20}Ne^+$, $^{22}Ne^+$, $N_2^+$, $^{14}N^{15}N^+$, $^{40}Ar^+$, and $^{136}Xe^+$ were obtained through data-fitting, while all other listed species in **Table 2** were respectively represented by one mass point. At *m/z* 28 and 29, counts were best fit with $N_2^+$ and $^{14}N^{15}N^+$ as the respective major species, which provided a $^{15}N/^{14}N$ ratio (5.40x10$^{-3}$ ± 1.77x10$^{-3}$), which is within the error limits to Earth values for $^{15}N/^{14}N$ (3.65x10$^{-3}$). Fits at *m/z* 29 yielded only ~1 count for $^{13}CO^+$ in one mass sweep, which translated to a theoretical upper limit of $CO^+$ in the LNMS of ≤10 counts and $C^{18}O^+$ being below the detection limit; as such, the respective abundances for $CO^+$ and $^{13}CO^+$ were not listed in **Table 2**. Mass positions with at least 1 mass sweep containing zero counts are underlined in **Table 2** and include $CO_2^+$, $^{38}Ar^+$, $NO^+$, and $^3He^+$ and/or $HD^+$.

*Table 2. List of background chemical species (m/z ≤136) in the LNMS measured at unknown pre-sampling altitudes as obtained from 3 complete mass sweeps and 1 incomplete spectrum (m/z ≤40), which are organized by calibrants (and related species), contaminants arising from terrestrial sources (e.g., gases attained during storage and residual chemicals from testing procedures), and chemicals potentially produced in situ within the mass analyzer; parent and related species are grouped together, potential isobars are grouped together, relative isotope abundances for Xe are provided in parentheses, $H_2^+$ and $O^+$ are cross-listed due to high relative abundances, and underlined species have a listed value of zero counts in ≥1 mass sweep in the archive data.*

| calibrants and related species | terrestrial contaminants | *in situ* species |
|---|---|---|
| $CH_4^+$, $^{13}CH_4^+$, $CH_3^+$, $CH_2^+$, $C^+$, $H_2^+$, and $H^+$ | $^4He^{2+}$ and/or $H_2^+$ | $HCN^+$, $CN^+$ |
| $^{124}Xe^+$ (0.008 ± 0.0001%) | $^3\underline{He}^+$ and/or $\underline{HD}^+$ | $C_2H_6^+$, $C_2H_5^+$, $C_2H_3^+$, $\underline{C_2H_2^+}$ |
| $^{128}Xe^+$ (0.023 ± 0.008%) | $^4He^+$, $^4He^{2+}$ | $\underline{NO^+}$ |
| $^{129}Xe^+$ (0.22 ± 0.07%) | $H_2O^+$, $OH^+$ | $SO_2^+$ |
| $^{130}Xe^+$ (0.063 ± 0.005%) | $^{20}Ne^+$ | |
| $^{131}Xe^+$ (0.20 ± 0.02%) | $^{22}Ne^+$ | |
| $^{132}Xe^+$, $^{132}Xe^{2+}$ (0.51 ± 0.05%) | $N_2^+$ | |
| $^{134}Xe^+$, $^{134}Xe^{2+}$ (4.2 ± 0.1%) | $^{14}N^{15}N^+$ | |
| $^{136}Xe^+$, $^{136}Xe^{2+}$, $^{136}Xe^{3+}$ (94.7 ± 0.2%) | $^{36}Ar^+$, $^{36}Ar^{2+}$ | |
| | $^{38}Ar^+$, $^{38}Ar^{2+}$ | |
| | $^{40}Ar^+$, $^{40}Ar^{2+}$ | |
| | $\underline{CO_2^+}$ | |
| | $^{86}Kr^+$ | |

For terrestrial contaminants, considerable counts arose from gases internal to the PV Large Probe (LP), which leaked into the LNMS during the ~3-4 month transit to Venus (Hoffman, Hodges, Donahue, et al., 1980, Hoffman, Oyama, et al., 1980) and potentially during pre-flight storage. As indicated in Oyama et al. (1980) and Hoffman, Oyama, et al. (1980), pressure in the PV Large Probe was adjusted to 1 atm during preflight procedures with $N_2$ containing 1% He (described further in **Section 3.2**). However, relative abundances in the pre-sampling data show an opposite relative enrichment with ~97% $^4He^+$ (19115 ± 1646 counts; n = 4 spectra, fitted) and ~3% $N_2^+$ (614 ± 90; n = 4 spectra, fitted). These trends are suggestive of selective diffusion of He into the LNMS through nanoscale leaks and/or the glass electrical feedthrough terminals (Hoffman, Hodges, Donahue, et al., 1980).

When considering the high abundances of $^4He^+$ in the pre-sampling data, the counts at *m/z* 2 (3810 ± 161, corrected; n = 4) and *m/z* 3 (6 ± 3; n = 2) are potentially inclusive of $^4He^{2+}$ and $^3He^+$, respectively. For this study, counts at *m/z* 2 were corrected for $H_2^+$ arising from $CH_4^+$ fragmentation using fitted counts for $CH_4^+$ and a value of 2.2% for the ratio of $H_2^+/CH_4^+$ (70 eV), as inferred from cross sections in Ward et al. (2011). For *m/z* 3, $^3He^+$ is a plausible assignment in addition to $HD^+$ and $H_3^+$. When assuming $^3He^+$ as the major species in the pre-sampling data, we obtain $^3He/^4He$ ratios (3.32 x $10^{-4}$ ± 1.62 x $10^{-4}$; n = 2) that are comparable to values (~$10^{-4}$) from select terrestrial sediments (Ozima et al., 1984, Takayanagi and Ozima, 1987) but are ~100-fold higher than terrestrial atmospheric values (Mishima et al., 2018). We posit that mass-dependent fractionation of He (Harrison et al., 2004) could have also occurred after restricted diffusion into the LNMS at the low temperatures during transit; however, the original helium isotope ratio and temperature of the transiting spacecraft are presently unknown.

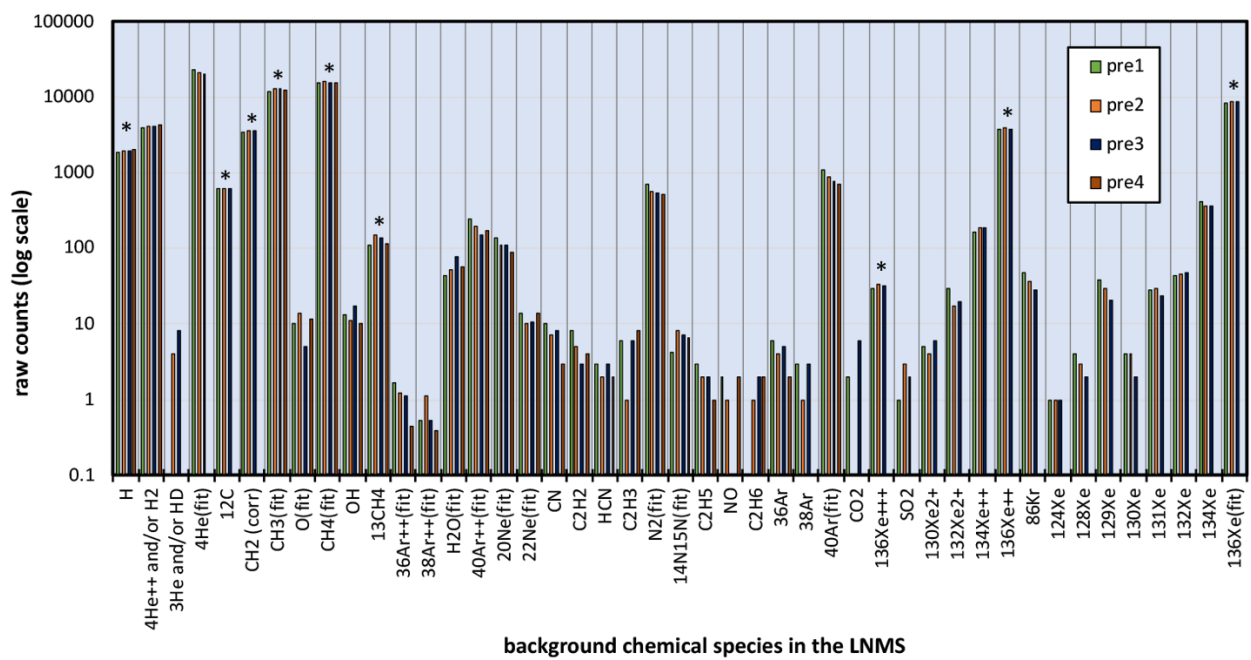

**Figure 2**. Inventory of background chemical species in the LNMS as obtained from fits to selected species (fit) or tracking of one mass point in the data; columns labeled as 'pre 1-4' represent the 4 consecutive mass sweeps in the pre-sampling data, and species used to measure repeatability in the LNMS are designated by an asterisk (*).

The pre-sampling data additionally harbor substantial counts for $^{40}Ar^+$ (930 ± 176; n = 4) and trace levels of $^{86}Kr$ (38 ± 10; n = 3), which respectively decrease by 37 and 41% in abundance across the sequential pre-sampling mass sweeps, which is suggestive of partial sequestering of the gases by the instrument getters. Relative isotope abundances for $^{36}Ar^+$ (0.5 ± 0.2%) and $^{40}Ar^+$ (99.3 ± 0.3%) are within error to terrestrial isotope ratios (Haynes, 2016) for Ar ($^{36}$Ar, 0.336%; $^{40}$Ar$^+$, 99.600%). However, $^{38}$Ar$^+$ (0.2 ± 0.1%) in the pre-sampling data is higher than terrestrial values ($^{38}$Ar$^+$, 0.063%), which is likely due to the unaccounted errors arising from the very low counts (1-3 counts) and single mass point representation of $^{38}$Ar$^+$. Counts for H$_2$O$^+$ (58 ± 18; n = 4) showed high scatter across the mass sweeps, but no clear trends towards loss; relative abundances for OH$^+$ as a fragmentation product (22.6 ± 5.3%, n = 4) matched NIST values (21%), while values for O$^+$, expected at ~2 counts (NIST, 3%), could not be reliably disentangled from isobaric CH$_4^+$.

The data also harbor potential counts for $^{22}$Ne$^+$ (11 ± 2; n = 4, fitted) at *m/z* 22. Counts arising from CO$_2^{2+}$ (*m/z* 22) were considered insignificant given the trace (2 and 6 counts) to negligent levels (0 counts in two spectra) of CO$_2^+$ (*m/z* 44). No other reasonable isobaric species were identified at *m/z* 22 for the closed LNMS. For $^{20}$Ne$^+$ in the pre-sampling data, counts at *m/z* 20 were disentangled from isobaric $^{40}$Ar$^{2+}$ using counts for $^{22}$Ne$^+$ and the terrestrial ratio for $^{22}$Ne/$^{20}$Ne (10%). We note that zero counts are observed at *m/z* 10 and 11 in the pre-sampling data, which negates detection of $^{20}$Ne$^{2+}$ and $^{22}$Ne$^{2+}$ and, in turn, the potential presence of $^{20}$Ne$^+$ and $^{22}$Ne$^+$. However, given the measured counts at *m/z* 22 (*e.g.*, $^{22}$Ne$^+$), mitigating explanations include rapid loss of trace $^{20}$Ne$^{2+}$ and $^{22}$Ne$^{2+}$ through collisions with $^{136}$Xe (Kahlert et al., 1983, Smith et al., 1980). Possible sources of Ne contamination include carryover from preflight LNMS testing, carryover from unreported leak testing procedures for the PVLP (*e.g.*, He leak testing (Bourcey et al., 2008)), and unaccounted leaks from the He-Ne lasers used in the cloud particle size spectrometer and nephelometer on the PVLP (Fimmel et al., 1995).

Very low counts (<10) are observed at mass positions consistent with SO$_2^+$, CO$_2^+$, NO$^+$, HCN$^+$, and CN$^+$, which are suggestive of trace contaminants (*e.g.*, SO$_2^+$ and CO$_2^+$) arising from ground-based controls (Hoffman, Hodges, Donahue, et al., 1980) and/or potential trace combination reactions (*e.g.,* NO$^+$, HCN$^+$, and CN$^+$) in the mass analyzer involving oxygen from H$_2$O, nitrogen from N$_2$, and carbon and hydrogen from CH$_4$. For NO$^+$, such assumptions provide a ratio of NO$^+$/N$_2^+$ of 0.3 ± 0.1%. Similarly, low counts (<10) for C$_2$H$_6^+$, C$_2$H$_5^+$, C$_2$H$_3^+$, and C$_2$H$_2^+$ are suggestive of trace combination-like reactions occurring after ionization and fragmentation of $CH_4$; among these species, $C_2H_2^+$ exhibits the maximum relative abundance of 0.03 ± 0.01% against $CH_4^+$. Together, these trends are suggestive of potential background reactions occurring in very low yield (~0.03-0.3%) under baseline operating conditions.

### *3.2 Internal Probe Pressure and Temperature*

We extracted trends from the literature regarding the changes in pressure and temperature during descent of the internal environment of the PV Large Probe, which protected instruments such as the LNMS and PVGC from the atmosphere through a titanium shell. As indicated in Seiff et al. (1995), no anomalous internal pressures or temperatures in the PVLP were noted through the descent. This indicates that nominal or expected temperature changes were experienced by the LNMS and PVGC throughout the descent. We note that the full internal pressure and temperature profiles and specifications of the internal temperature sensors are not yet available in the released PV materials.

Nevertheless, as summarized in Oyama et al. (1980), Hoffman, Oyama, et al. (1980), and Seiff et al. (1995), pressure within the PVLP was (1) adjusted to 1 atm during preflight procedures with $N_2$ containing 1% He for the purposes of testing for leaks in the hermetically sealed probe and preventing breakdown of the electrical leads of the LNMS during operation, (2) re-adjusted to ~1.3 bar (~1.3 atm, as expressed in Oyama et al. (1980), and 19.0 psia, as expressed in Seiff et al. (1995)) with additional $N_2$ just prior to atmospheric entry at Venus via a pre-programmed sequence aimed at compensating for the expected (though ultimately negligible) losses in pressure during transit, (3) measured at 1.8 bar (1.8 atm, 27.27 psia) upon touchdown at the surface, and (4) associated with internal temperature increases of 0.8-1.14 ˚C at each sampling interval.

These total observations are suggestive of increases in gas abundances by ~1.6 moles within the internal probe environment along with final internal temperature increases of ~50-60 ˚C upon landing on Venus' surface. These calculations rely on an initial internal probe pressure of 1.3 bar, a *presumed* initial probe temperature of -10-0 ˚C, 50 *presumed* internal temperature measurements across the descent, a final internal pressure of 1.8 bar, and a *presumed* void volume (~180 L) for the PVLP of 10%.

Equivalently, our calculations indicate that ≥1.5 moles of gas could have been released by the PVGC, which discharged carrier gas (He) – and the discarded atmospheric samples – into the probe interior during operation (Oyama et al., 1980). These calculations assume a constant flow rate of 35 mL/min (17 bar, 18.3 ˚C) (Oyama et al., 1980) and total operation time of 60 min, which includes the purge step prior to atmospheric entry (coarsely estimated at ~5 min) and the full ~55 min descent profile (~1 min from ~200-67 km, and ~54 min from ~67-0 km). Thus, the LNMS was likely subjected to increasing partial pressures of He throughout the descent.

### *3.3 Statistical Insights into the LNMS Data*

Statistical uncertainties for ion counts in the LNMS were obtained by tracking the internal calibrants – methane ($CH_4$), xenon (Xe), and associated products – in the pre-sampling data. In the original LNMS investigations, uncertainties were estimated using the Poisson distribution (square root of the counts), which is independent of the LNMS operations and yields unrealistically low errors for higher count species (*e.g.*, ≤3% for counts of ≥1000). Instead, we extracted in situ measures of error by tracking 9 different internal ionization and fragmentation standards (**Table 1**) across the pre-sampling mass sweeps. For $CH_4$, ionization and fragmentation from 4 mass sweeps yielded averaged counts and standard deviations for $^{13}CH_4^+$ (127 ± 18 raw counts), $CH_4^+$ (15,707 ± 355; fitted), $CH_3^+$ (12,613 ± 454; fitted), $CH_2^+$ (3571 ± 110), $C^+$ (613 ± 9), and $H^+$ (1952 ± 69). Tracking of $^{136}Xe$ ionization across 3 mass sweeps yielded averaged counts and standard deviations for $^{136}Xe^+$ (8671 ± 137; fitted), $^{136}Xe^{2+}$ (3840 ± 111) and $^{136}Xe^{3+}$ (32 ± 2); we note that data at masses higher than *m/z* 40 were not recorded/transmitted in one pre-sampling spectrum.

The combined standard deviations from the measures ranged from 1.4-3.6% for the fitted species, with nearly identical ranges 1.5-3.5% for species represented by one mass point with counts of >600. These trends are indicative repeatable mass spectral measurements, when assuming each pre-sampling spectra to be equivalent.

Therefore, to estimate uncertainty in the descent data, we applied a minimum error of 3.6% to all fitted and higher count species (counts >600), where errors from the regressions for fitted species were propagated if higher. Conversely, the standard deviations ranged from 6.3-14.2% for species represented by one mass point with counts of <130, which is indicative of considerable error arising from *unaccountable* shifts in the mass scale (peak centers) and/or changes in the FWHM (peak widths). Hence, in the descent data, we conservatively applied a minimum error of 14.2% to all single mass point species with low counts (counts <600).

### 3.4 Peak Shapes in the LNMS

Performance of the LNMS across the descent was assessed by comparison of the experimental masses (peak centers) and FWHM (peak widths) for the internal peak standards (**Table 1**). The single-species fits for $CH_3^+$, $^{40}Ar^+$, and $^{136}Xe^+$ (at 70 eV from 51.3, 13.3, and 0.9 km) are displayed in **Figs. 3A-C**, while example fits to mass clusters at *m/z* 16 and 18 are provided in **Figs. 3D-E**, which together illustrate the high-quality regressions using a Gauss function. Displayed in **Figs. 4A-C** are vertical profiles for the mass error ($\Delta(m/z)$) for $CH_3^+$, $^{40}Ar^+$, and $^{136}Xe^+$ across the pre-sampling (rounded boxes) and descent (64.2 to 0.9 km) profiles, where $\Delta(m/z)$ represents the difference between the experimental and exact mass (Brenton and Godfrey, 2010).

**Table 3.** List of fitted parameters for the internal peak standards ($CH_3^+$, $CO^+$, $N_2^+$, $^{40}Ar^+$, and $^{136}Xe^+$) including the averaged experimental masses, range in mass errors, averaged full width half maxima (FWHM), and range in FWHM, as calculated across 64.2-0.9 km for $^{40}Ar^+$ and $^{136}Xe^+$, and 64.2-0.2 km for $CH_3^+$, $CO^+$, and $N_2^+$; mass errors ($\Delta(m/z)$) are the difference between the experimental mass and exact mass, uncertainties represent the standard deviation, and counts at 135.742 amu at 2.2 km were treated as an outlier and removed for fits at m/z 136.

| internal standards | averaged mass (*m/z*) | mass error range | averaged FWHM (*m/z*) | FWHM range (*m/z*) |
|---|---|---|---|---|
| $CH_3^+$ | 15.023 ± 0.001 | -0.003 – 0.001 | 0.018 ± 0.001 | 0.015 – 0.020 |
| $CO^+$ | 27.995 ± 0.001 | -0.017 – 0.001 | 0.040 ± 0.003 | 0.033 – 0.042 |
| $N_2^+$ | 28.006 ± 0.003 | -0.019 – 0.009 | 0.041 ± 0.004 | 0.033 – 0.042 |
| $^{40}Ar^+$ | 39.967 ± 0.002 | -0.001 – 0.012 | 0.065 ± 0.005 | 0.058 – 0.081 |
| $^{136}Xe^+$ | 135.934 ± 0.027 | -0.046 – 0.086 | 0.277 ± 0.031 | 0.214 – 0.327 |

Mass errors across the descent profile are minimal and show no systematic errors or influences from the external pressures and temperature; however, mass accuracies decrease at higher *m/z*. Listed in **Table 3** are the averaged experimental masses from the descent profile including the ranges in the mass error. Together, these parameters are indicative of the LNMS exhibiting appreciable mass accuracies (Brenton and Godfrey, 2010) during atmospheric sampling with values of 0.0005 Da (32 ppm) for $CH_3^+$, 0.0046 Da (114 ppm) for $^{40}Ar^+$, and 0.0277 Da (204 ppm) for $^{136}Xe^+$ – these values are based on the absolute mass errors and are expressed using 4 decimal places (Habibzadeh and Habibzadeh, 2015) to yield the reported 3 decimal places in the LNMS mass scale (Hoffman, Hodges, Donahue, et al., 1980).

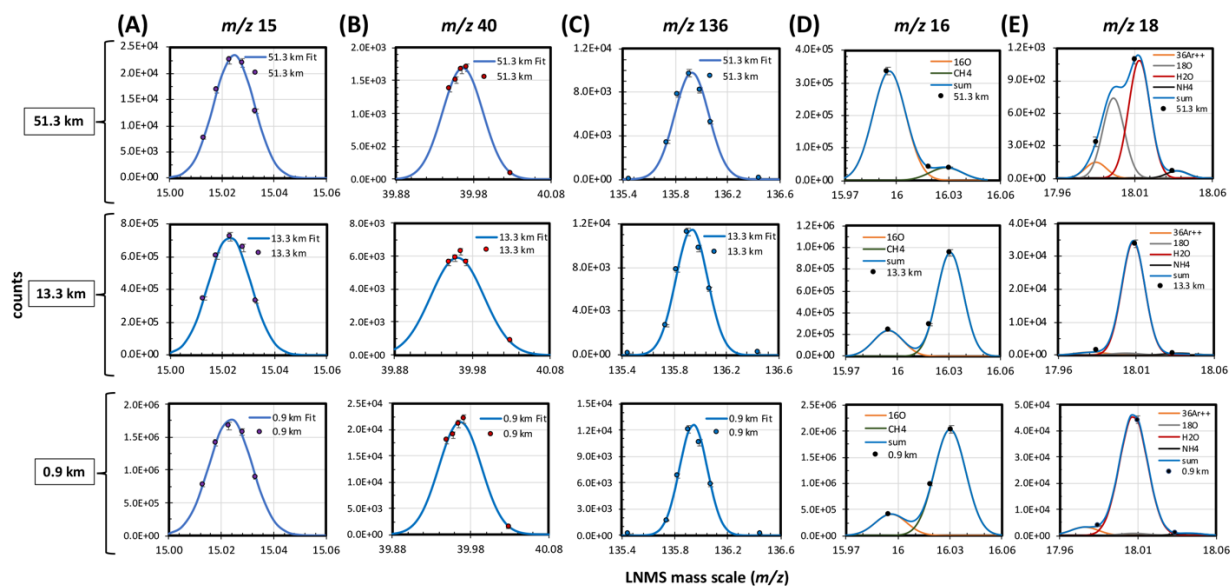

**Fig. 3**. Data fitting using Gauss functions for selected internal peak standards and mass clusters from spectra obtained at 51.3 km (top row), 13.3 km (middle row), and 0.9 km (bottom row), where the internal peak standards include (A) $CH_3^+$ (*m/z* 15), (B) $^{40}Ar^+$ (*m/z* 40), and (C) $^{136}Xe^+$ (*m/z* 136), and mass clusters include (D) $^{16}O^+$ and $CH_4^+$ (*m/z* 16), and (E) $^{36}Ar^{2+}$, $^{18}O^+$, $H_2O^+$, and $NH_4^+$ or $NH_2D$ (*m/z* 18); blue line denotes the minimized fit across the included species (legend), mass scale is denoted as *m/z*, y-axis is commensurately scaled for clarity, and error bars represent the standard deviation of the measure.

Displayed in **Figs. 4A-C** are trends for the FWHM for $CH_3^+$, $^{40}Ar^+$, and $^{136}Xe^+$ across the descent profile (64.2-0.9 km), while the averaged FWHM and ranges in the FWHM are listed in **Table 3**. Standard deviations for the FWHM are minimal across the descent yet *increase* slightly with *m/z* (~8% for $CH_3^+$, ~9% for $^{40}Ar^+$, and ~11% for $^{136}Xe^+$). Oppositely, standard deviations for the FWHM in the pre-sampling data *decrease* with *m/z* (~7% for $CH_3^+$, ~1% for $^{40}Ar^+$, and 0.3% for $^{136}Xe^+$). These trends suggest that atmospheric sampling yields increased variances in peak width in a *m/z*-dependent manner.

The effects of peak height (counts) on the FWHM were addressed by tracking $CH_3^+$ and $^{40}Ar^+$ across the descent profile. As inferred from **Fig. 3A**, counts for $CH_3^+$ unexpectedly increase by ~1000-fold across the descent (as described further in the **Discussion**), yet no major changes to the FWHM are observed; however, the lowest FWHM values for $CH_3^+$ are obtained during the main clog. Similarly, as inferred from **Figs. 3B** and **4B**, the ~40-fold increases in the $^{40}Ar^+$ abundances yield no systematic changes to the FWHM; however, the peak widening near ~13 km (**Fig. 4B**) and potential peak sharpening at 0.9 km (**Fig. 3B**) are suggestive of the presence of isobaric species such as $Ne_2^+$ and $MgO^+$.

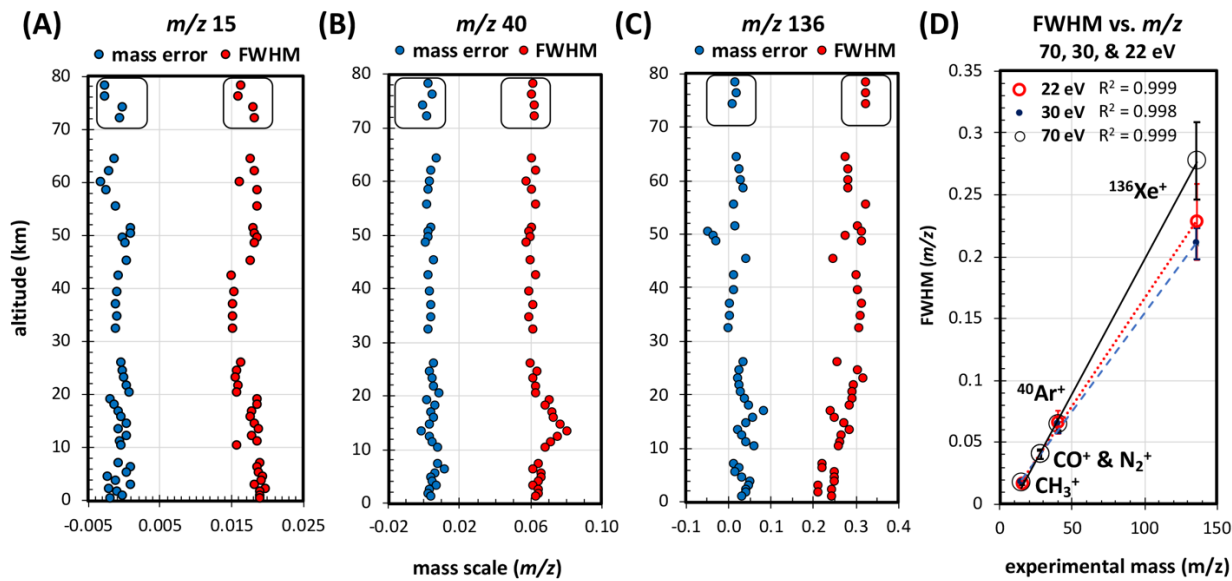

**Fig. 4**. Trends in peak width (FWHM) and mass error ($\Delta m/z$) for the internal peak standards. Altitude profiles (64.2-0.2 km) for FWHM and mass error obtained at 70 eV are provided for (A) $CH_3^+$ (*m/z* 15), (B) $^{40}Ar^+$ (*m/z* 40), and (C) $^{136}Xe^+$ (*m/z* 136), and include the pre-sampling mass sweeps (rounded boxes), which are nominally listed between 78-72 km for clarity. Linear regressions across the averaged FWHM and averaged experimental masses respectively obtained at 70, 30, and 22 eV are provided in (D) for $CH_3^+$, $CO^+$ (*m/z* 28), $N_2^+$ (*m/z* 28), $^{40}Ar^+$, and $^{136}Xe^+$; $R^2$ values represent the quality of linear regressions across the averaged values.

Together, the trends for FWHM and Δ($m/z$) reveal *no anomalies* in the operations of the mass analyzer during the clog (*e.g.*, noise, significant and unexplained fluctuations, contradictory trends, *etc.*), after recovery in the lower atmosphere, or near the surface at the higher atmospheric pressures and temperature. Instead, the trends, as displayed in **Fig. 4D**, reveal linear correlations ($R^2$ = 0.998 ± 0.002, 38 spectra) between FWHM and $m/z$ for the internal peak standards ($CH_3^+$, $CO^+$, $N_2^+$, $^{40}Ar^+$, and $^{136}Xe^+$) in all descent spectra obtained at 70 eV. Linear correlations are also obtained at 30 eV ($R^2$ = 0.996 ± 0.006, 3 spectra) and 22 eV ($R^2$ = 0.995 ± 0.008, 3 spectra) between FWHM and the $m/z$ for $CH_3^+$, $^{40}Ar^+$, $^{136}Xe^+$ (**Fig. 4D**). Comparisons across electron energies show that the FWHM for $CH_3^+$ and $^{40}Ar^+$ are respectively identical across 22-70 eV, while the FWHM for $^{136}Xe^+$ increases slightly at 70 eV.

Regarding the mass resolving power of the LNMS ($m/\Delta m$; Nic et al. (2006)), our data fitting for $m/z$ 16 at 49.4 km yields an *in situ* value of 453 with a 9.4% separation (between $CH_4^+$ and $O^+$), which is nearly identical to the mass resolving power reported in Hoffman, Hodges, Wright, et al. (1980) of 440 with a 9% separation (between $CH_4^+$ and $O^+$), which we presume to be a ground-based measurement. Additionally, for the low mass channel, we obtain a resolving power of ~35 using the FWHM for $^4He^+$ at $m/z$ 4, as compared to the mass resolving power of ~15 with a <1% separation inferred from Hoffman, Hodges, Wright, et al. (1980).

### *3.5 Ionization and Transmission in the LNMS*
#### 3.5.1 Tracking $^{136}Xe^{n+}$ and $^{40}Ar^{n+}$

Ionization and transmission efficiencies (*i.e.*, fraction of ions that pass through the ion source and reach the detector) in the LNMS were assessed by tracking $^{136}Xe^{n+}$ and $^{40}Ar^{n+}$ in the pre-sampling and descent (64.2-0.9 km) profiles. Displayed in **Fig. 5A** are the full vertical profiles for $^{136}Xe^+$ ($m/z$ 136, fitted), $^{136}Xe^{2+}$ ($m/z$ 68.0), and $^{136}Xe^{3+}$ ($m/z$ 45.3). Due to the IRMC sample injection, spikes in the counts are observed at 45.2 km for $^{136}Xe^+$ and $^{136}Xe^{2+}$, yet no commensurate spike is observed for $^{136}Xe^{3+}$, potentially due to collisional loss. For $^{136}Xe^+$, $^{136}Xe^{2+}$, and $^{136}Xe^{3+}$, baseline counts respectively increase by ~1.2, ~1.5, and ~1.7-fold after atmospheric intake, as indicated by comparison of counts from the clouds (64.2-51.3 km) and pre-sampling data. After closure of Valve 1, under the conditions of the lower atmosphere (24.4-0.9 km), baseline counts for $^{136}Xe^+$, $^{136}Xe^{2+}$, and $^{136}Xe^{3+}$ respectively increase by ~1.1, ~1.2, and ~1.8-fold, when compared to the cloud data. These combined trends suggest that increasing mass flow intake rates along the descent yield increases in the transmission efficiencies at decreasing *m/z*.

When statistically tracked across the descent (64.2-51.6 km and 24.4-0.9 km; 70 eV), similar standard deviations (~11%) are obtained from the averaged counts for $^{136}Xe^+$ (10971 ± 1227; fitted) and $^{136}Xe^{2+}$ (6345 ± 698), while much higher errors (46%) are exhibited by $^{136}Xe^{3+}$ (78 ± 36). For $^{136}Xe^{3+}$, the total trends are suggestive of a competing isobaric species at *m/z* 45.3. Similar assessments are drawn from the low eV spectra (30 and 22 eV). As inferred from cross sections in Stephan and Märk (1984), ionization of Xe to yield $Xe^{3+}$ is not expected at ≤35 eV. Nevertheless, counts at *m/z* 45.3 are readily observed in the 30 and 22 eV spectra obtained from the middle clouds (7 counts, 54 km; 10 counts, 52.7 km) and lower atmosphere (30 counts, 9.1 km; 16 counts, 8.2 km).

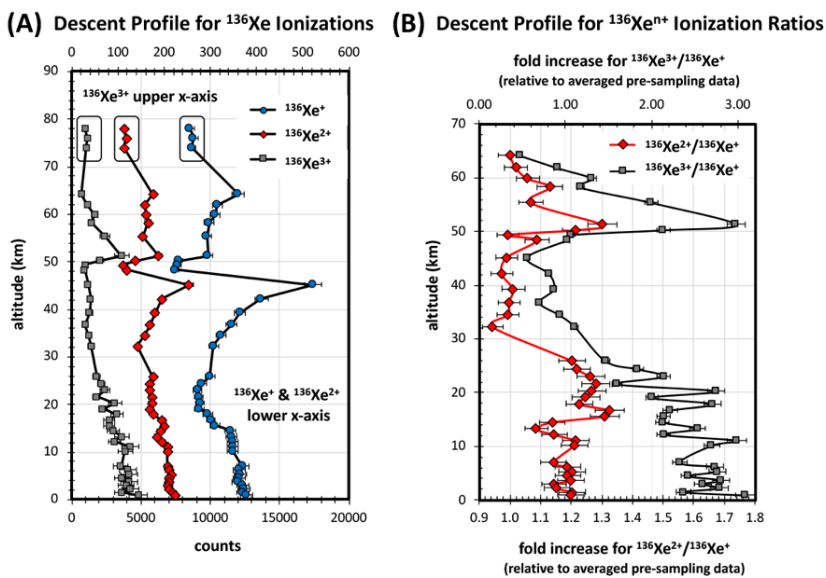

**Fig. 5**. Altitude profiles (64.2-0.9 km) for (A) $^{136}Xe^+$ (blue circles, lower x-axis), $^{136}Xe^{2+}$ (red diamonds, lower x-axis), and $^{136}Xe^{3+}$ (gray squares, upper x-axis), inclusive of the pre-sampling data, which are nominally listed between 78-72 km for clarity (rounded boxes), and (B) the ratios of $^{136}Xe^{2+}/^{136}Xe^+$ (red diamonds) and $^{136}Xe^{3+}/^{136}Xe^+$ (gray squares), which are expressed as the change or fold increase relative to the pre-sampling data; propagated errors are displayed.

In contrast, zero counts are obtained during the main clog (30.1 and 28.1 km), at low eV (30 and 22 eV), where atmospheric intake was limited. Yet, the nominal mass sweeps continued during the main clog, inclusive of in-flight corrections using $^{136}Xe^{n+}$. These observations suggest that the isobar at *m/z* 45.3 arises from outside the mass analyzer. A possible isobar for $^{136}Xe^{3+}$ (*m/z* 45.302) is $^{181}Ta^{4+}$. In theory, tantalum ions $Ta^{n+}$, inclusive of $Ta^{4+}$, could be released or dissociated from the surfaces of the inlet tubes, which potentially reacted with atmospheric $CO_2$ during the initial mass sweeps (see **Section 3.6**). Mass positions for $Ta^{2+}$ and $Ta^{3+}$ were not measured by the LNMS, and $Ta^{3+}$ is potentially *not* a source of $Ta^{4+}$ at 30 and 22 eV, given the calculated ionization energy of 36 eV for $Ta^{3+}$ (Carlson et al., 1970). No other reasonable isobars were identified at *m/z* 45.3.

**Table 4.** *Comparison of the LNMS ionization ratios (%) for $Xe^{2+}/Xe^+$, $Xe^{3+}/Xe^+$, and $Ar^{2+}/Ar^+$ to values extracted from the literature (superscripts) and NIST database. Averaged LNMS values (70 eV) were calculated using relative abundances of $^{136}Xe^{n+}$ and $^{40}Ar^{n+}$ from the pre-sampling and descent (64.1-51.3 km and 24.4-0.9 km) data. Ground based ionization ratios (70 eV) were extracted from Fox (1959), Fox (1960), and the NIST database using measures of relative abundance (e.g., intensity, counts, or current). Ionization ratios (70-75 eV) were also calculated using partial cross sections from Stephan and Märk (1984), Straub et al. (1995), and Rejoub et al. (2002).*

| Ionization Ratios | Relative Abundances | | | | Cross Sections |
|---|---|---|---|---|---|
| | LNMS (pre) | LNMS (descent) | Literature | NIST | Literature |
| $Xe^{2+}/Xe^+$ | 44 ± 1% | 58 ± 4% | [a]35% | ~19% | [b]7.7%, [c]8.3% |
| $Xe^{3+}/Xe^+$ | 0.37 ± 0.02% | 0.81 ± 0.24% | [a]0.4% | – | [b]0.1% (75 eV) |
| $Ar^{2+}/Ar^+$ | 22 ± 2% | 25 ± 3% | [d]13% | ~17% | [e]5.5% |

[a] Fox (1959)
[b] Stephan and Märk (1984)
[c] Rejoub et al. (2002)
[d] Fox (1960)
[e] Straub et al. (1995)

### 3.5.2 Tracking the $^{136}Xe^{n+}/^{136}Xe^+$ Ratios

Impacts of the descent conditions (*e.g.*, clouds, main clog, and lower atmosphere) on the ionization yields and transmission efficiencies were also assessed using the $^{136}Xe^{2+}/^{136}Xe^+$ ratio (**Table 4**). Across the descent, excluding the main clog (~50-25 km), the averaged $^{136}Xe^{2+}/^{136}Xe^+$ ratio (56 ± 5%) exhibits a standard deviation of ~9% across 38 spectra, which is suggestive of moderate reproducibility with respect to single and double ionization of the parent species and transmission of the respective ions towards to the detector (at *m/z* 136 and 68) – when the inlet/s to the LNMS were open to the atmosphere.

When compared to values from the pre-sampling data, the ionization ratio of

$^{136}Xe^{2+}/^{136}Xe^{+}$ also reveals a dependence upon atmospheric intake into the LNMS. As shown in **Fig. 5b**, the $^{136}Xe^{2+}/^{136}Xe^{+}$ ratios, after commencement of sampling at 64.2 km (0.12 bar), and when relative to the pre-sampling data, (1) increase ~1.3-fold by 51.3 km (0.88 bar), (2) decrease back to pre-sampling values during the main clog (~50-25 km), (3) increase again to ~1.3-fold by 16.7 km (28.9 bar), and (4) then slightly decrease to ~1.2-fold between 15.6 to 0.9 km. As described in **Section 3.5.1**, the observed trends relate to enhancements in transmission efficiency at *m/z* 68 for $^{136}Xe^{2+}$, when compared to *m/z* 136, and suggest that mass flow intake rates impact ion transmission in a *m/z*-dependent manner. In addition, below 15.6 km, the relative $^{136}Xe^{2+}/^{136}Xe^{+}$ ratios show intermittent decreases of ~1.1-fold (relative to the pre-sampling data) across 14.5-12.2 km, 10.2-6.2 km, and 3.0-2.2 km, which are suggestive of impacts to intake rates.

The maximum relative increases of ~1.3-fold in the $^{136}Xe^{2+}/^{136}Xe^{+}$ ratios (at 51.3 and 16.7 km) across the descent also reveal that optimal transmission efficiencies (at *m/z* 68) were obtained when total counts in the LNMS were ~$6\times10^6$, which was achieved at 51.3 km ($6.0\times10^6$ total counts) and 16.7 km ($5.6\times10^6$ total counts). Our analysis (**Section 3.9**) of control data for $CO_2$ from Hoffman, Hodges, Wright, et al. (1980) shows that the LNMS yields respectively similar counts ($1.5\times10^6$ and $1.1\times10^6$ counts) at external pressures at 51.3 km (0.88 bar) and 16.7 km (28.9 bar). Thus, the total mass flow rates of intake during descent were likely similar at 51.3 km (0.88 bar) and 16.7 km (28.9 bar) due to closure of the secondary inlet with the higher flow rate via Valve 1 at ~1.6 bar (~47 km) and opening of the VCV.

### 3.5.3 Comparison to Literature Values

Compared in **Table 4** are the ionization ratios for $Xe^{n+}/Xe^{+}$ and $Ar^{2+}/Ar^{+}$ at 70 eV obtained from terrestrial measures and the LNMS (pre-sampling and descent data). For the LNMS, ionization ratios for $^{136}Xe^{2+}$ in the pre-sampling data are ~1.3-fold higher than ratios calculated from relative abundances (70 eV) extracted from Fox (1959), ~2.3-fold higher than ratios calculated from relative abundances in the NIST database (70 eV), and ~5-6-fold higher than ratios inferred from cross sections (70 eV) in Rejoub et al. (2002) and Stephan and Märk (1984). Ionization ratios for $Xe^{3+}$ in the pre-sampling data are effectively equivalent to values (70 eV) reported in Fox (1959), yet are ~4-fold higher than values inferred from cross sections (75 eV) in Stephan and Märk (1984). Similar trends are observed for $^{40}Ar^{2+}$ ($Ar^{2+}/Ar^+$) in the pre-sampling data, where ionization ratios are ~2.3-fold higher than values (70 eV) extracted from Fox (1960), ~1.7-fold higher than NIST values (70 eV), and ~5.4-fold higher than values inferred from cross sections (70 eV) in Straub et al. (1995).

These pre-sampling data trends are suggestive of significant enhancements by the electron multipliers for multiple charged ions. As shown in Fox (1960), relative abundances of double and triple-charged noble gases increase by ~2–5-fold with use of an electron multiplier (Dumont 6291, 10 stage, dynode). As inferred from Fox (1960) and Schram et al. (1966), multiple charged ions exhibit higher resultant kinetic energies upon striking the first dynode of the multiplier due to acceleration through the magnetic field and difference in potential between the ion source and first dynode.

### 3.5.4 Potential for $^4He^{2+}$ in the Pre-Sampling Data

Given the high background counts for $^4He^+$ in the pre-sampling data (19115 ± 1646), we present the assignment of $^4He^{2+}$ at *m/z* 2, rather than solely $H_2^+$. In support are the described enhancements at the electron multiplier for multiple charged ions, and the dedicated electron multiplier in the LNMS that served as a low mass channel. When assuming $^4He^{2+}$ as the major species at *m/z* 2, we obtain roughly comparable ionization ratios for $^4He^{2+}/^4He^+$ from the pre-sampling data (18 ± 2%) *and* IRMC sequence at 45.2 km (~12%). During the IRMC sequence, counts at *m/z* 2 increase by ~2-fold between 48.5 and 45.2 km, which is consistent with the mild increases across the single and double charged noble gases in the data immediately following the IRMC injection. Per our understanding, enrichment in $H_2^+$ is not an expected result of the IRMC sequence since reactive gases were scrubbed during the gas enrichment procedure – as confirmed by the ~250-fold *decrease* in counts for $CO_2^+$ between 48.5 and 45.2 km.

To yield $He^{2+}$, critical constraints for the ionization include (1) a measured appearance energy of ~78-80 eV for electron ionization of He to directly yield $He^{2+}$ (Denifl et al., 2002, Fox, 1959), which is slightly higher than the 70 eV used in the LNMS, (2) a measurable cross section (2.03x10$^{-18}$ cm$^2$) at 70 eV for the ionization of He$^+$ (Dolder et al., 1961), which indicates that He$^+$ could be ionized to yield He$^{2+}$ in the LNMS, (3) a calculated energy of ~54 eV to ionize He$^+$ to He$^{2+}$ (Carlson et al., 1970, Kramida et al., 2022), which is well under the 70 eV used in the LNMS, and (4) the suggestion in Dolder et al. (1961) that collisions between He$^+$ and residual gases in a mass spectrometer also yield He$^{2+}$.

Thus, in the context of the LNMS, the high ionization ratios for $^4$He$^{2+}$/$^4$He$^+$ likely necessitate the sequential ionization of $^4$He to yield $^4$He$^{2+}$ and/or higher electron energies stemming from unidentified influences on the LNMS electronics package. We posit that sequential ionization to yield $^4$He$^{2+}$ was favored due to the multi-point diffusion of He into the LNMS during transit and potentially during operation, which potentially increased the residence times of He and He$^+$ near or within the electron and ion sources.

### *3.6 Fragmentation in the LNMS*

Review of methane (CH$_4$) fragmentation from the pre-sampling and descent data, which are summarized in **Table 5**, provided additional insights into the baseline performance of the LNMS. For CH$_4$$^+$ and CH$_3$$^+$, respective counts were obtained by data fitting as described, while counts for CH$_2$$^+$, C$^+$, H$_2$$^+$, and H$^+$ were represented each by one mass point. For CH$_4$$^+$, fitted counts were corrected for $^{13}$CH$_3$$^+$ using fitted counts for CH$_3$$^+$ and the $^{13}$C/$^{12}$C ratio from CH$_4$ in the pre-sampling data (0.80 ± 0.11), which had high error (14.2%) due to the single mass point representation for $^{13}$CH$_4$$^+$ and low counts (127 ± 18 raw counts). For CH$_2$$^+$, counts at *m/z* 14 across the pre-sampling and descent profiles were corrected for $^{14}$N$^+$ using fitted counts for N$_2$$^+$ and the ratio of *m/z* 14/28 (0.16) from the NIST standard for N$_2$; resultant negative values from the corrections (*e.g.*, when N$_2$$^+$ >> CH$_2$$^+$) were discarded, as were enriched values at 42.2 and 23.0 km (~240%), which stem from scatter in the data or an unaccounted (though temporary) source of counts at *m/z* 14 during the main clog.

From the pre-sampling data (**Table 5**), our analyses provide relative abundances (CH$_4$$^+$,

base peak) for $CH_3^+$, $CH_2^+$, and $C^+$ that are *essentially identical* to values inferred from cross sections (70 eV) in Liu and Shemansky (2006), Sigaud and Montenegro (2015), and Ward et al. (2011), and similar to ranges reported in the MassBank EU and NIST databases. These trends are indicative of nominal or expected ionization and fragmentation yields in the LNMS just prior to atmospheric sampling.

When considering $H^+$ as a product of $CH_4^+$ fragmentation (**Table 5**), relative abundances in the pre-sampling data are ~2-fold *lower* than values inferred from cross sections in Ward et al. (2011) and ~2-fold *higher* than relative abundances (~6-7%) extracted from spectra (50-100 eV) in Adamczyk et al. (1966), Dibeler and Mohler (1950), and Gluch et al. (2003). For $H^+$, these trends are suggestive of the LNMS harboring reasonable but sub-optimal transmission efficiencies at *m/z* 1. When considering $H_2^+$ as a product of $CH_4^+$ fragmentation, and if assigning all counts at *m/z* 2 to $H_2^+$, a relative abundance of ~26% is obtained, which is ~10-30-fold higher than the ~1-2% inferred from cross sections in Ward et al. (2011) and Gluch et al. (2003). Thus, $CH_4$ fragmentation to yield $H_2^+$ does not significantly contribute to *m/z* 2 in the LNMS.

*Table 5. Comparison of the methane ($CH_4$) fragmentation profile from the LNMS to literature sources and the NIST and MassBank EU databases. Abundances of $CH_3^+$, $CH_2^+$, $C^+$, and $H^+$ are expressed as a percent (%) relative to $CH_4^+$. For the pre-sampling data (pre) from the LNMS (70 eV), relative abundances were averaged across 3-4 spectra, while relative abundances across the descent profile (descent) were averaged across 25 spectra (70 eV) from the clouds (58.3-51.3 km) and lower atmosphere (24.4-0.9 km); values are listed using 2 significant figures for simplicity and standard deviations are provided. Relative abundances were also extracted from the NIST and MassBankEU databases (70 eV), and inferred using cross sections (70 eV) from Liu and Shemansky (2006), Sigaud and Montenegro (2015),* and *Ward et al. (2011).*

| Fragment | Relative Abundances | | | Cross Sections |
|---|---|---|---|---|
| | LNMS (pre) | LNMS (descent) | NIST and MassBankEU | Literature[a] |
| $CH_3^+$ | 80 ± 2% | 79 ± 7% | 83-89% | 81-83% |
| $CH_2^+$ | 23 ± 1% | 20 ± 2% | 15-21% | 19-21% |
| $C^+$ | 3.9 ± 0.1% | – | 3-5% | 3-4% |
| $H^+$ | 12 ± 1% | 11 ± 2% | – | 22% |

[a] Liu and Shemansky (2006), Sigaud and Montenegro (2015), Ward et al. (2011)

Across the descent profile (64.2-0.9 km), the counts for $CH_4^+$ unexpectedly increase by ~3-

log from ~$10^3$ to ~$10^6$ counts, as inferred from **Fig. 3A**, while counts for $CH_3^+$, $CH_2^+$, $C^+$, and $H^+$ commensurately increase to yield relative abundances that are statistically equivalent to those from the pre-sampling data, though with higher relative errors (**Table 5**). For the $CH_3^+$ fragmentation profile (79.4 ± 7.0%), the relatively high error (~9%) across the descent is mostly representative of variances between 64.2-50.3 km (or 71.8 ± 8.4%), where relative abundances between 49.4-0.9 km show less error (81.5 ± 4.4%). These trends, irrespective of the relative increases in counts, are supportive of nominal fragmentation behavior of the LNMS mass analyzer through the descent.

*3.7 Organics Contamination*

Vacuum sealant and benzene were cited as potential terrestrial contaminants in the original investigations (Donahue et al., 1981, Hoffman, Hodges, Donahue, et al., 1980). However, by our estimates, benzene was not a major terrestrial or Venus contaminant in the LNMS, despite a large spike in counts between 16.7-10.2 km (30,208 counts at 13.3 km). When respectively attributing raw counts at *m/z* 78 (78.053 amu) and *m/z* 79 (78.924 amu) to benzene ($C_6H_6$) and the peak shoulder of the $^{13}C$-isotopomer of benzene ($C_5(^{13}C)H_6^+$), we obtain a $^{13}C/^{12}C$ ratio ranging from $1.3 \times 10^{-4}$ to $3.0 \times 10^{-4}$ at ≤24.4 km. This range in apparent $^{13}C/^{12}C$ ratios for benzene in the LNMS is ~30-80-fold lower than the terrestrial $^{13}C/^{12}C$ value ($1.1 \times 10^{-2}$; (Haynes, 2016)), terrestrial $^{13}C/^{12}C$ value obtained from benzene mass spectra ($1.1 \times 10^{-2}$; MassBank EU), $^{13}C/^{12}C$ value for atmospheric $CO_2$ from Venus ($1.260 \times 10^{-2} \pm 0.072 \times 10^{-2}$; **Section 3.8**), and $^{13}C/^{12}C$ value for methane from the LNMS data ($0.89 \times 10^{-2} \pm 0.14 \times 10^{-2}$; pre-sampling data).

Benzene, therefore, was not a major species in the LNMS. Nevertheless, explanations for the benzene being depleted in $^{13}C$ to yield $\delta^{13}C$ values of -988 to -973‰ (however unlikely) include organic contamination arising from (1) $^{12}C$-enriched adhesives, sealants, greases, and/or cleaning solvents – which are likely prohibitively costly to synthesize or purchase – or (2) *in situ* formation of $^{12}C$-enriched benzene via a massive – and likely implausible – mass-dependent fractionation of atomic carbon, which could theoretically arise from unknown origins and be released from the spectrometer walls. Alternatively, the counts at *m/z* 78 could be representative of isobars such as SO(CH$_3$)$_2^+$ (dimethyl sulfoxide, DMSO), P$_2$O$^+$, and/or AsH$_3^+$ (arsine) and speculate that (1) DMSO may arise from reactions within the mass analyzer between CH$_3^+$ (internal calibrant) and SO gas arising from dissociated sulfates or sulfuric acid at the LNMS inlet, (2) P$_2$O$^+$ may arise from the dissociation of phosphorus anhydrides (P$_x$O$_y$) at the LNMS inlet, and (3) AsH$_3^+$ may potentially be a by-product of reactions (Doak et al., 2000, Scott et al., 1989) at the inlet between water and trapped arsenides (*e.g.*, Fe$_2$As and Na$_3$As), which are potentially stable at Venus conditions (Lewis and Fegley, 1982).

Nevertheless, in this study, we accounted for a *maximum* potential for contamination at all mass positions corresponding to benzene, which included the parent ion (C$_6$H$_6^+$), the $^{13}$C-isotopomer (C$_5$($^{13}$C)H$_6^+$), and several fragmented products (C$_6$H$_5^+$, C$_6$H$_3^+$, C$_5$H$_2^+$, C$_5$H$_1^+$, C$_4$H$_4^+$, C$_4$H$_3^+$, C$_4$H$_2^+$, C$_3$H$_4^+$, C$_3$H$_2^+$, C$_3$H$_1^+$, C$_2$H$_4^+$, C$_2$H$_3^+$, C$_2$H$_2^+$, and CH$_3^+$). Counts were corrected by (1) assigning counts at *m/z* 79 to C$_5$($^{13}$C)H$_6^+$, (2) obtaining the expected counts for C$_6$H$_6^+$ and associated fragments using the relative ratios from the benzene spectral reference in the MassBank EU database, and (3) subtracting the expected counts from each respective mass position in the LNMS data.

### *3.8 Carbon Dioxide Ionization and Fragmentation*

Displayed in **Figs. 6A-C** are altitude profiles for the products of CO$_2$ ionization and fragmentation inclusive of CO$^{18}$O$^+$, $^{13}$CO$_2^+$, C$^{18}$O$^+$, $^{13}$CO$^+$, CO$^+$, $^{13}$CO$_2^{2+}$, CO$_2^{2+}$, O$^+$, and C$^+$. Percent abundances are listed in **Table 6** and expressed relative to CO$_2^+$. Averages and standard deviations were calculated across the clouds (58.3-51.3 km) and lower atmosphere (24.4-0.9 km), as well as aggregated across the descent (58.3-51.3 km, 24.4-0.9 km). Data from the main clog were omitted. Potential sources of unaccounted error(s) include the single mass point tracking of CO$_2^+$, CO$^{18}$O$^+$, $^{13}$CO$_2^+$, and $^{13}$CO$_2^{2+}$. For C$^+$, counts arising from CH$_4^+$ were estimated using the apparent C$^+$/CH$_4^+$ ratio from the pre-sampling data (3.9%), as detailed in **Section 3.5**, and fitted counts for CH$_4^+$.

Across the descent (**Table 6**), relative abundances for the single and double ionized isotopologues of $CO_2$ ($CO^{18}O^+$, $^{13}CO_2^+$, $^{13}CO_2^{2+}$, and $CO_2^{2+}$) show minimal error and are equivalent across the clouds and lower atmosphere.  Hence, single and double ionization of the parent ions after entry in the lower atmosphere (24.4-0.9 km) were not impacted by the closure of Valve 1 (~47 km), the temporary obstruction of the LNMS inlets across ~50-25 km, ensuing influx of chemical species during the main clog (~50-25 km), or opening of the VCV as atmospheric pressure increased.

Across the descent (58.3-51.3 and 24.4-0.9 km), averaged relative abundances for all species are slightly higher than NIST values, which is contrary to the results obtained for $CH_4^+$ fragmentation.  For $CO_2^+$, these results are suggestive of enhanced transmission efficiencies in the design of the LNMS (when compared to the NIST standard), along with slightly higher carbon and oxygen isotope abundances for Venus (**Section 3.8**).  For $CO_2^{2+}$, data fitting at *m/z* 22 yields relative abundances that are ~1.4-fold higher than NIST values and ~3-fold higher than relative abundances inferred from cross sections in King and Price (2008) and Straub et al. (1996), which is suggestive of the presence of isobaric $^{20}Ne^+$, along with enhancements at the electron multiplier for the multiply charged $CO_2^{2+}$.

Tracking of $CO^+$ and $C^+$ reveals trends that are consistent with dissociative losses of $CO_2$ (and/or $CO_2^+$) between 64.2-59.9 km (**Figs. 6A** and **6B**).  At 64.2 km, the first altitude of atmospheric sampling, relative abundances of $CO^+$ (75%) and $C^+$ (214%) are very high, while counts for $CO_2^+$ (5504 raw counts) are unexpectedly low in the initial mass sweep.  Relative abundances for $CO^+$ (30%) and $C^+$ (38%) then significantly decrease by 59.9 km and ultimately stabilize between 58.3-51.3 km (23 ± 3% $CO^+$, and 19% $C^+$).  Concomitantly, counts for $CO_2^+$ increase by ~300-fold across 64.2-51.3 km with near-expected values being reached between 58.3-51.3 km ($\geq 10^5$ raw counts) via linear incremental gains of ~$1.5 \times 10^5$ counts km$^{-1}$ between 59.9-51.3 km.  In parallel, as summarized in **Table 6** and **Fig. 6C,** counts for $O^+$ track with $CO_2^+$ to yield stable relative abundances across the clouds (58.3-51.3 km) with no observed enrichments upon atmospheric sampling.

Hence, data from the initial mass sweeps between 64.2-59.9 km are consistent with the dissociation of $CO_2$ or $CO_2^+$ upon sampling. In the LNMS, the dissociation was observed as relative enrichments in $CO^+$ and $C^+$, along with an absence of enriched $O^+$ due to potential loss of oxygen to the interior spectrometer surfaces (*e.g.*, inlets, ion source cavity, or magnetic sector). In Hoffman, Hodges, Donahue, et al. (1980), loss of $CO_2$ was attributed to the cleaned nature of the spectrometer walls; however, $CO_2$ dissociation was not discussed nor mentioned. We note that Ta surfaces, as described in Belov et al. (1978), can promote $CO_2$ dissociation to yield desorbed CO and surface-retained O. Hence, the interior surfaces of the Ta inlet tubes could serve as sites for $CO_2$ dissociation and sinks for oxygen.

*Table 6. Comparison of the altitude profile for carbon dioxide ($CO_2$) fragmentation in the LNMS to literature sources and the NIST database. Abundances of $CO^{18}O^+$, $^{13}CO_2^+$, $C^{18}O^+$, $^{13}CO^+$, $^{13}CO_2^{2+}$, $CO_2^{2+}$, $O^+$, and $C^+$ are expressed as a percent (%) relative to $CO_2^+$ and listed using 2 significant figures for simplicity. Averaged relative abundances and standard deviations were obtained for the clouds (3 spectra, 70 eV, 58.3-51.3 km), lower atmosphere (LA; 22 spectra, 70 eV, 24.4-0.9 km), and across the descent profile (25 spectra, 70 eV, 58.3-51.3 and 24.4-0.9 km); however, the cloud data contain only one measure for $C^+$ in the clouds between 58.3-51.3 km, as the missing data were likely not transmitted during flight. Relative abundances were also extracted from the NIST database (70 eV) and inferred using cross sections (75 and 70 eV, respectively) from King and Price (2008) and Straub et al. (1996).*

| Ion/Fragment | m/z | Relative Abundances | | | | Cross Sections |
|---|---|---|---|---|---|---|
| | | LNMS (clouds) | LNMS (LA) | LNMS (descent) | NIST | Literature |
| $CO^{18}O^+$ | 46 | 0.45 ± 0.01% | 0.44 ± 0.03% | 0.44 ± 0.03% | 0.29% | – |
| $^{13}CO_2^+$ | 45 | 1.3 ± 0.1% | 1.3 ± 0.1% | 1.3 ± 0.1% | 1.1% | – |
| $C^{18}O^+$ | 30 | 0.059 ± 0.002% | 0.036 ± 0.009% | 0.034 ± 0.012% | – | – |
| $^{13}CO^+$ | 29 | 0.36 ± 0.02% | 0.20 ± 0.04% | 0.22 ± 0.06% | – | – |
| $CO^+$ | 28 | 23 ± 3% | 11 ± 2% | 12 ± 5% | 9.7% | [a]18%, [b]19% |
| $^{13}CO_2^{2+}$ | 22.5 | 0.032 ± 0.002% | 0.033 ± 0.002% | 0.033 ± 0.002% | – | – |
| $CO_2^{2+}$ | 22 | 2.7 ± 0.1% | 2.6 ± 0.1% | 2.6 ± 0.1% | 1.8% | [a]0.75%, [b]0.93% |
| $O^+$ | 16 | 20 ± 1% | 19 ± 1% | 19 ± 1% | 9.5% | [a]23%, [b]24% |
| $C^+$ | 12 | 19% | 17 ± 1% | 15 ± 2% | 8.6% | [a]12%, [b]13% |

[a] Straub et al. (1996); 70 eV
[b] King and Price (2008); 75 eV

However, as inferred from Hoffman, Hodges, Wright, et al. (1980), sample degradation by the Ta inlets was minimized by passivation (with $H_2SO_4$) to yield a tantalum oxide inner surface.

Further, per our understanding of the pre-flight studies, the LNMS was routinely subjected to high temperature bakeout procedures with no apparent adverse effects to the ensuing measurements, which is suggestive of the cleaned spectrometer walls harboring minimal degradative potential. Thus, the unexpectedly low abundances of $CO_2$ measured by the LNMS between 64.2-59.9 km remain to be fully explained.

Listed in **Table 6** are the averaged relative abundances for the $CO_2$ fragmentation products, as calculated across 58.3-51.3 km after equilibration of the $CO_2$ counts between 64.2-59.9 km. Counts for $CO^+$ between 58.3-51.3 km were assumed to arise predominantly from $CO_2^+$ fragmentation due to the low atmospheric abundances of CO (Oyama et al., 1980). Tracking of $CO^+$ and $^{13}CO^+$, and $C^{18}O^+$ as fragment ions reveals ~1.6-2-fold decreases in relative abundances between the clouds (58.3-51.3 km) and lower atmosphere (24.4-0.9 km), which is contrary to the trends for the isotopologous parent ions that showed no difference. When compared to terrestrial values, relative abundances for $CO^+$ from the clouds (23 ± 3%) are ~2-fold higher than NIST values and ~1.3-higher than values inferred from cross sections (King and Price, 2008, Straub et al., 1996).

For the cloud data, the $CO^+$ trends are suggestive of optimal transmission efficiencies prior to closure of Valve 1; unaccounted errors include the single mass point analyses. For the lower atmospheric data, the ~2-fold lower relative abundances for $CO^+$ (when compared to the clouds) are suggestive of losses to the chemical getter (G1), which was gradually exposed to the ion source cavity via the VCV, as atmospheric pressure increased. Further, the trends for $CO^+$ between 24.4-0.9 km suggest that opening of the VCV yields roughly constant percent losses in $CO^+$, presumably due to proportionate increases between transmission rates through the mass analyzer and retention rates by the getter (G1).

In sharp contrast, opening of the VCV yields no discernable impacts on the relative abundances of $O^+$ and $C^+$ (**Table 6**) – and by extension $CO_2^+$. Rather, relative abundances for $O^+$ and $C^+$ are identical in the clouds and lower atmosphere. When compared to terrestrial values, relative abundances for $O^+$ and $C^+$ are ~2-fold higher than NIST values but are respectively ~1.2-fold lower and ~1.2-fold higher than values inferred from cross sections. These trends are suggestive of optimal transmission efficiencies for $O^+$ and $C^+$ in the LNMS, when compared to the respective cross sections and after considering for unaccounted errors arising from the single mass point analyses. These trends also suggest that transmission towards the getter and/or retention by the getter are ion-dependent, as only $CO^+$ and isotopologues are noticeably impacted by the VCV and the changes in transmission pathways, while no major impacts are observed for $CO_2^{n+}$ (and isotopologues), $O^+$, and $C^+$.

In Hoffman, Hodges, Donahue, et al. (1980), the counts at *m/z* 32 were attributed to $O_2^+$ arising from fragmentation of $CO_2^+$. In our understanding, control studies with the pre-flight LNMS yield a $O_2^+/CO_2^+$ ratio of $1.94 \times 10^{-4}$ at 100 eV (obtained via personal review by M. Way of unreleased archived PV documentation at NASA ARC). In our analysis of the LNMS data, best fits at *m/z* 32, when including $^{32}S^+$ as the sole isobar, yield substantial counts for $O_2^+$ in the clouds (~310 counts, 51.3 km) and lower atmosphere (~1260 counts, 0.9 km). Our ensuing $O_2^+/CO_2^+$ ratios for the clouds ($1.75 \times 10^{-4}$; 51.3 km) are consistent with the pre-flight LNMS control values; however, values in the lower atmosphere ($5.8 \times 10^{-4}$; 0.9 km) are ~3-fold higher.

In comparison, high-resolution studies in King and Price (2008) regarding the electron ionization of $CO_2$ (30-200 eV), which measured relative cross sections as low as $~1 \times 10^{-5}$, show no evidence of $O_2^+$ formation. Hence, these contradictory results for $O_2^+$ suggest that design of the LNMS mass analyzer (*e.g.*, miniature size, electron multiplier, etc.) may have promoted very low yield combination-type reactions. This assessment is consistent with trends from the pre-sampling data, where $C_2H_2^+$ is formed from $CH_4$ in relative yield of ~0.03%, similar to the relative yield of $O_2^+$ formation (~0.06%; 0.9 km). Additionally, $O_2^+$ could be produced from other oxygen containing particles and gases (*e.g.*, water) in the lower atmosphere through low yield combination-type reactions.

Lastly, in the lower atmosphere at ~15 ± 2 km, slight and temporary increases are observed (at 15.6 km) in the relative abundances for CO$^+$ (~1.9-fold; $m/z$ 28), $^{13}$CO$^+$ (~1.8-fold; $m/z$ 29), C$^{18}$O$^+$ (~2.4-fold; $m/z$ 30), and CO$^{18}$O$^+$ (~1.2-fold; $m/z$ 46). The differential increases across species are suggestive of (A) atmospheric sources for CO$^+$ ($m/z$ 28) at ~15 ± 2 km, such as oxycarbon species (*e.g.*, CO$_3$ and CO$_3^{2-}$), and (B) the presence of isobaric species for C$^{18}$O$^+$ ($m/z$ 30) and CO$^{18}$O$^+$ ($m/z$ 46). At $m/z$ 29, relevant isobars for $^{13}$CO$^+$ include CHO$^+$. At $m/z$ 30, relevant isobars for C$^{18}$O$^+$ include CH$_2$O$^+$ and NO$^+$. At $m/z$ 46, relevant isobars for CO$^{18}$O$^+$ include CH$_2$O$_2^+$. These combined trends are suggestive of formic acid as a potential parent species (parent ion, CH$_2$O$_2^+$ ($m/z$ 46); base peak, CH$_2$O$^+$ ($m/z$ 30); and fragment, CHO$^+$ ($m/z$ 29)). At $m/z$ 30, potential sources of NO$^+$ include nitric oxide gas, fragmentation of nitric acid (HNO$_3$), or dissociation of nitrate (NO$_3^-$) at the LNMS inlets.

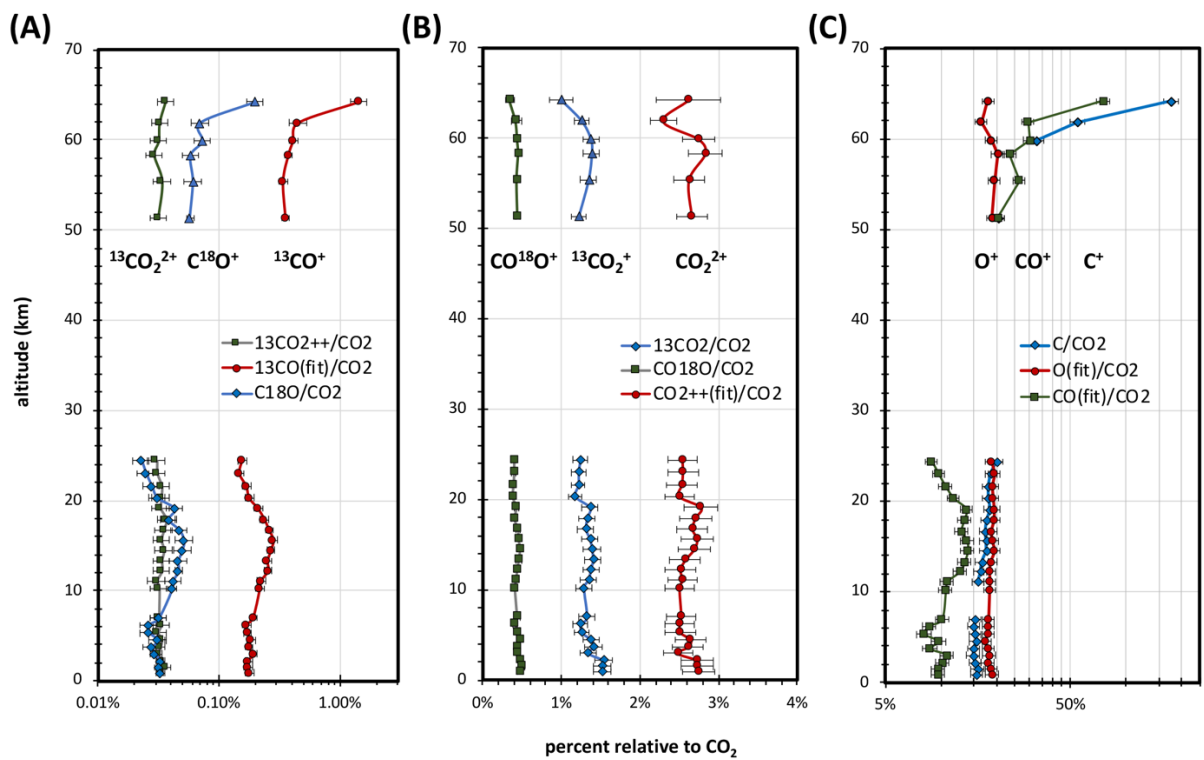

**Fig. 6**. Altitude profiles (64.2-0.9 km) for ionized and fragmented products of CO$_2$ inclusive of CO$^{18}$O$^+$, $^{13}$CO$_2^+$, C$^{18}$O$^+$, $^{13}$CO$^+$, CO$^+$, $^{13}$CO$_2^{2+}$, CO$_2^{2+}$, O$^+$, and C$^+$, which are expressed as the relative percent to CO$_2^+$ and organized for clarity by those with percent values in the lower atmosphere of (A) ≤1.1%, which include $^{13}$CO$_2^{2+}$, C$^{18}$O$^+$, and $^{13}$CO$^+$ (log scale), (B) ≤4%, which included CO$^{18}$O$^+$, $^{13}$CO$_2^+$, and CO$_2^{2+}$ (linear scale), and (C) 5-65%, which included CO$^+$ (fitted values), O$^+$, and C$^+$ (log scale displayed in units of five); error bars represent the propagated error.

*3.9 Carbon and Oxygen Isotope Ratios*

Tracking $CO_2^+$, $CO_2^{2+}$, and $CO^+$ along with the isotopologues of $^{13}CO_2^{2+}$, $^{13}CO_2^+$, and $CO^{18}O^+$ yielded $^{13}C/^{12}C$ and $^{18}O/^{16}O$ ratios across the Venusian atmosphere. For all reported isotope ratios, standard deviations were larger than the propagated errors. For $^{13}C/^{12}C$, the altitude profiles for $^{13}CO_2^+/CO_2^+$ (1.338 x $10^{-2}$ ± 0.096 x $10^{-2}$), $^{13}CO_2^{2+}/CO_2^{2+}$ (1.260 x $10^{-2}$ ± 0.072 x $10^{-2}$), and $^{13}CO^+/CO^+$ (1.487 x $10^{-2}$ ± 0.111 x $10^{-2}$) yield a combined average of 1.363 x $10^{-2}$ ± 0.132 x $10^{-2}$ (~10% standard deviation) across 75 spectra (58.3-51.3 km and 24.4-0.9 km). However, a general trend of $^{13}CO_2^{2+}/CO_2^{2+}$ < $^{13}CO_2^+/CO_2^+$ < $^{13}CO^+/CO^+$ is respectively observed across the profiles for the clouds (58.3-51.3 km), lower atmosphere (24.4-0.9 km), and descent (58.3-51.3 km and 24.4-0.9 km). These trends are suggestive of the Venus $^{13}C/^{12}C$ value being best represented by the $^{13}CO_2^{2+}/CO_2^{2+}$ ratio (1.260 x $10^{-2}$ ± 0.072 x $10^{-2}$; 5.7% standard deviation, 25 spectra), with the higher ratio for $^{13}CO^+/CO^+$ (1.487 x $10^{-2}$ ± 0.111 x $10^{-2}$) being suggestive of interfering isobaric species. In addition, the $^{13}CO_2^+/CO_2^+$ and $^{13}CO^+/CO^+$ ratios respectively increase by ~13 and 11% between the altitudes of 3.0 and 0.9 km (data not shown), which is suggestive of interfering isobaric species or near-surface changes in the Venus $^{13}C/^{12}C$ value. Isotope effects in the mass analyzer are unlikely explanations for the trends since studies in Inoue et al. (1964) indicate negligible enrichments in the $^{13}C/^{12}C$ ratio for $C^+$ (1.008 ± 0.002) after electron ionization and fragmentation of $CO_2$. Potential isobars against $^{13}CO_2^+$ (*m/z* 45) and $^{13}CO^+$ (*m/z* 29) include $CHO_2^+$ (*m/z* 45) and $CHO^+$ (*m/z* 29).

For the $^{18}O/^{16}O$ ratio, the altitude profiles for $CO^{18}O^+/CO_2^+$ (2.195 x $10^{-3}$ ± 0.156 x $10^{-3}$) and $C^{18}O^+/CO^+$ (2.558 x $10^{-3}$ ± 0.283 x $10^{-3}$) yield a combined average of 2.376 x $10^{-3}$ ± 0.291 x $10^{-3}$ (~12% standard deviation) across 75 spectra (58.3-51.3 km and 24.4-0.9 km). However, a general trend of $CO^{18}O^+/CO_2^+$ < $C^{18}O^+/CO^+$ is respectively observed across the profiles for the clouds (58.3-51.3 km), lower atmosphere (24.4-0.9 km), and descent (58.3-51.3 km and 24.4-0.9 km). These trends are suggestive of the Venus $^{18}O/^{16}O$ ratio being best represented by the $CO^{18}O^+/CO_2^+$ ratio (2.195 x $10^{-3}$ ± 0.156 x $10^{-3}$; 7.1% standard deviation, 25 spectra), with the higher ratio for $C^{18}O^+/CO^+$ (2.558 x $10^{-3}$ ± 0.283 x $10^{-3}$) being suggestive of interfering isobaric species. In addition, between 3.0-0.9 km, the differential increases of ~9 and 32% in the respective $CO^{18}O^+/CO_2^+$ and $C^{18}O^+/CO_2^+$ ratios (data not shown) are again suggestive of interfering isobaric species. Potential isobars against $CO^{18}O^+$ (*m/z* 46) include $CH_2O_2^+$, while isobars against $C^{18}O^+$ (*m/z* 30) include $CH_2O^+$ and $NO^+$.

### *3.10 Normalization of LNMS Counts*

Raw counts were normalized to $^{136}Xe^{2+}$ to account for the impacts of the descent on ionization and transmission efficiencies. This strategy contrasts with the analyses in Hoffman, Hodges, Donahue, et al. (1980) where $CO_2$ was not normalized, and from Donahue and Hodges (1993) where $CH_4^+$ was normalized to $^{136}Xe^+$.

We normalized $CO_2$ (*m/z* 43.990) to $^{136}Xe^{2+}$ (*m/z* 67.954) for the reasons provided in the following list.

(1) $CO_2^+$ (*m/z* 44) and $^{136}Xe^{2+}$ (*m/z* 68) have closer masses when compared to $^{136}Xe^+$ (*m/z* 136).

(2) The transmission coefficients of $CO_2^+$ (~0.97) and $^{136}Xe^{2+}$ (~0.46) are more similar when compared to $^{136}Xe^+$ (~0.12), as measured in Donahue and Hodges (1992).

(3) Increases in mass flow intake yield larger impacts to the transmission efficiencies at *m/z* 68 when compared to *m/z* 136, as inferred from tracking the $^{136}Xe^{n+}$ descent profiles.

(4) The relative abundances of $Xe^{2+}$ are significantly less susceptible to influences from the electron multiplier when compared to $Xe^{3+}$ (*m/z* 45.302), as demonstrated in Fox (1959).

(5) Data were not normalized to $^{136}Xe^{3+}$ since relative abundances are likely skewed due to isobaric species.

To account for opening of the primary and secondary inlets at ~64.2 km (clouds) and closure of the secondary inlet via Valve 1 at ~47 km (lower atmosphere), we separately normalized data from the clouds and lower atmosphere. For the cloud data (64.2-51.3 km), our analyses show that counts for $^{136}Xe^{2+}$ and $^{136}Xe^+$ immediately spike after opening of the inlets (64.2 km) and reach stable values across 61.9-58.3 km. When compared to the pre-sampling data, averaged counts across 61.9-58.3 km for $^{136}Xe^{2+}$ and $^{136}Xe^+$ are ~1.4- and 1.2-fold higher, respectively, and show minimal variance ($^{136}Xe^{2+}$, 5376 ± 128, ~2% error, n = 3; $^{136}Xe^+$, 10269 ± 296, fitted, ~3% error, n = 3).  Hence, the cloud data were normalized to counts of $^{136}Xe^{2+}$ at 61.9 km or when the mass analyzer equilibrated after opening of the inlets.  Data from the lower atmosphere (24.4-0.9 km) were normalized to averaged values for $^{136}Xe^{2+}$ (5674 ± 74, ~1% error, n = 3) between 24.4-21.6 km to account for closure of Valve 1 and the ~1.5-fold increase in baseline values when compared to the pre-sampling data.

For data from the main clog, $^{136}Xe^{n+}$ could not be used for normalization due to the IRMC sequence, which substantially increased counts for $^{136}Xe^+$ and $^{136}Xe^{2+}$ between 45.2-25.9 km.  As a coarse alternative, counts from the main clog were normalized to the ratio of $^{134}Xe^{2+}/^{136}Xe^{2+}$ from 49.4 km.  In control studies, normalization of the lower atmospheric data (≤24.4 km) to the $^{134}Xe^{2+}/^{136}Xe^{2+}$ ratio yielded poor matches to expected values.

### 3.11 Conversion to $CO_2$ Density

Normalized counts for $CO_2$ were converted to units of density (kg m$^{-3}$) using calibration curves assembled from published data (Hoffman, Hodges and Duerksen, 1979, Oyama et al., 1980, Seiff et al., 1985).  Construction of the calibration curves involved extraction of control data from Hoffman, Hodges and Duerksen (1979), which described a Venus-relevant pressure-temperature profile for $CO_2$ in the LNMS, and were predicated upon conversion of the published data into units of counts (LNMS output) from amps (detector output).

As outlined in **Fig. 7A**, the ratio of counts per amp for the LNMS before and after closure of Valve 1 were *estimated* by (1) converting the volumetric percent values for $CO_2$ at 51.6 and 21.6 km from Oyama et al. (1980) into mass percent values, (2) converting the mass percent values for $CO_2$ at 51.6 km (0.698 bar) and 21.6 km (17.8 bar) into densities (kg m$^{-3}$) using atmospheric data from Seiff et al. (1985), (3) converting the densities for $CO_2$ at 51.6 and 21.6 km into units of detector output (amps) using the pressure-temperature profile for $CO_2$ from Hoffman, Hodges and Duerksen (1979) (**Fig. 7B-C**), and (4) obtaining the ratio of counts/amp by dividing the LNMS counts for $CO_2$ at 51.3 and 21.6 km by the *expected* amps for $CO_2$ at 51.6 and 21.6 km, respectively.

Calibration curves were then obtained by re-expressing the $CO_2$ density profiles from Hoffman, Hodges and Duerksen (1979) into units of counts (from amps), where data obtained *before* and *after* the closure of Valve 1 were converted using the counts/amp ratios at 51.6 and 21.6 km, respectively. In turn, we used the calibration curves to separately convert normalized counts for $CO_2^+$ into densities (kg m$^{-3}$) for data from the clouds and lower atmosphere, or *before* and *after* closure of Valve 1, respectively.

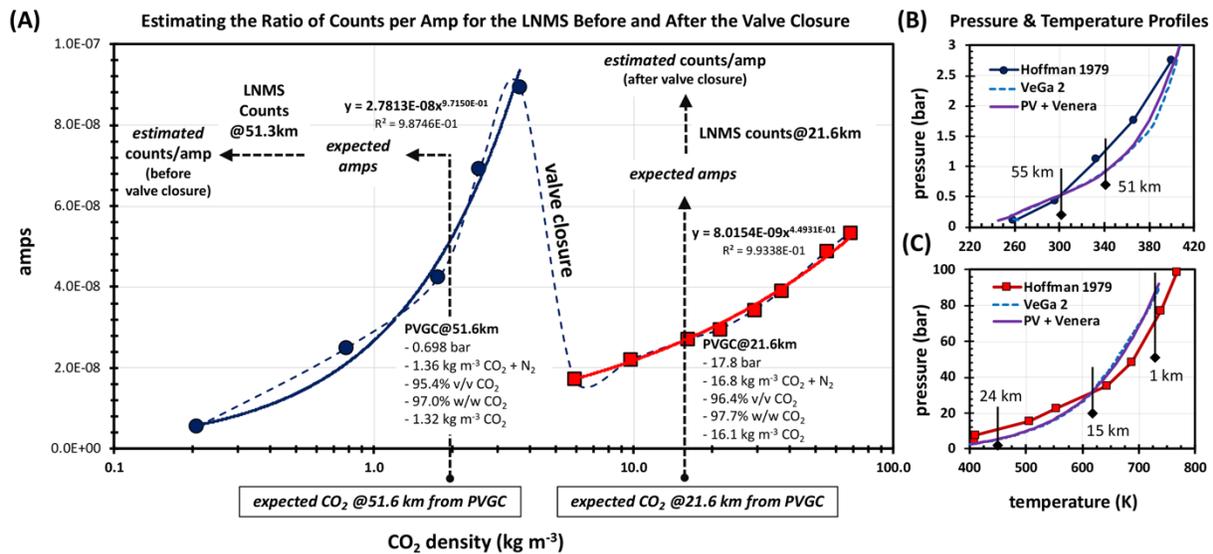

**Fig. 7**. Overview of data used to convert counts to densities for $CO_2$. (A) Diagram outlining the estimation of counts/amp using data extracted from Hoffman, Hodges and Duerksen (1979), which was expressed as $CO_2$ densities, and fitted using power functions ($R^2$ 0.987 and 0.993); counts/amp were obtained before and after the valve closure by sequential conversions of volumetric $CO_2$ measures from the PVGC at 51.6 and 21.6 km to $CO_2$ densities (using mass percent and atmospheric density), and $CO_2$ densities to *expected* amps (dotted arrows) using the power functions, followed by estimation of counts/amp by division of the normalized $CO_2$ counts in the LNMS at 51.3 and 21.6 km by the respective *expected* amps. (B and C) Pressure and temperature profiles for the $CO_2$ profile in Hoffman along with in situ measures from data collected by the VeGa 2 (blue dashed line) and combined Pioneer Venus and VeNeRa (purple line) spacecraft for conditions related to the (B) to the clouds (~64-45 km) and (C) lower atmosphere (~45-0 km); demarcations for the altitudes of 55, 51, 24, 15, and 1 km are provided.

Uncertainties for these conversions were estimated using the standard errors of the y-estimate from the calibration curves, which was 15.3% for data obtained *before* closure of Valve 1 (64.2-48.4 km), and 5.9% for data obtained *after* closure of Valve 1 (45.2-0.9 km). Additional sources of error include an assumption of constant counts/amp in each $CO_2$ calibration curve, along with the pressure and temperature profile used in Hoffman, Hodges and Duerksen (1979), which slightly differs from measured values from the VeGa 2, Pioneer Venus, and VeNeRa spacecraft (**Figs. 7B-C**) (Linkin et al., 1986, Seiff et al., 1985).

Qualitatively, comparison of the pressure and temperature profiles suggests that conversion of counts at 51.3 km (see demarcations for 55 and 51 km in **Fig. 7B**) may yield higher than expected densities due to the higher presumed pressures used in Hoffman, Hodges and Duerksen (1979) between 320-380 K. Conversely, conversion of counts at ≤15 km (see demarcations for 24, 15 and 1 km in **Fig. 7C**) may yield slightly lower than expected densities due to the lower presumed pressures used in Hoffman, Hodges and Duerksen (1979) between 650-750 K.

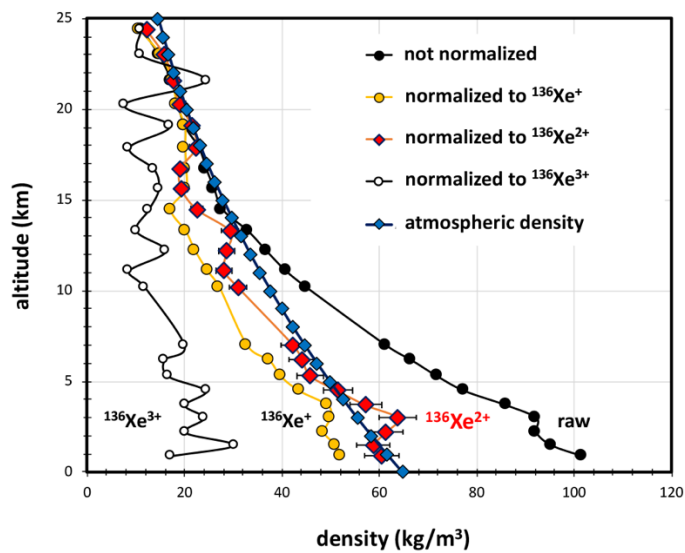

**Fig. 8**. Altitude profiles (64.2-0.9 km) for $CO_2$ before (black filled circles) and after normalization to $^{136}Xe^+$ (yellow circles), $^{136}Xe^{2+}$ (red diamonds), $^{136}Xe^{3+}$ (black empty circles), along with the bulk atmospheric density profile (blue diamonds), where error bars represent uncertainty from the model; as noted in **Section 3.10**, $^{136}Xe^{3+}$ does not provide a good normalization, likely due to isobaric species, whereas $^{136}Xe^{2+}$ and $^{136}Xe^+$ yield closer matches to the atmospheric density profile.

Displayed in **Fig. 8** are the $CO_2$ density profiles obtained after normalization to $^{136}Xe^+$, $^{136}Xe^{2+}$, and $^{136}Xe^{3+}$. The $CO_2$ altitude profiles are also compared to bulk atmospheric densities calculated from data in Seiff et al. (1985). When considering expected values of ~97% w/w $CO_2$, normalization to $^{136}Xe^{2+}$ provided the best relative match to atmospheric densities. Conversion of raw counts without normalization yielded substantially larger densities (≤160%) at ≤10 km; thus, highlighting the necessity for normalization to account for changes in transmission efficiencies towards the surface. Normalizations to $^{136}Xe^+$ provided reasonable matches between 24.4-19.1 km; however, densities are ≤40% of expected values below ~17 km; thus, highlighting the necessity to account for the *m/z*-dependent impacts to ion transmission as a result of intake. Normalization to $^{136}Xe^{3+}$ yielded very poor matches, as densities were ~25-70% lower than expected across 64.2-0.9 km, which is consistent with $^{136}Xe^{3+}$ being skewed by isobaric species.

### 3.12 Altitude Profiles for $CO_2$ and Atmospheric Density

Displayed in **Fig. 9** are the altitude profiles for $CO_2$ densities (normalized to $^{136}Xe^{2+}$) and Venus atmospheric densities across the clouds and lower atmosphere. Potential uncertainties from the conversion of $CO_2^+$ counts to densities (standard error) were estimated using the standard error of the y-estimate from the calibration curves (15.6% for data from the clouds, and 5.9% for data from the lower atmosphere). For averaged densities, the uncertainties (propagated standard error of the mean) were obtained by propagating the standard errors and dividing by the square root of the sample size (n). Variances in the measures were estimated using the standard deviation.

The LNMS data between 64.2-59.9 km (**Fig. 9A**; red diamonds) reveal lower than expected $CO_2$ densities (0.0011-0.20 kg m$^{-3}$) during the initial sampling sweeps, when assuming an expected abundance of ~97% w/w $CO_2$ (green squares). Note, densities between 64.2-61.1 km serve as estimates since the respective counts are far below the bounds of the calibration curve (**Fig. 7**).

After equilibration of $CO_2$ counts, $CO_2$ densities in the clouds at 58.3 km (0.38 ± 0.06 kg m$^{-3}$), 55.4 km (0.80 ± 0.12 kg m$^{-3}$) and 51.3 km (1.41 ± 0.22 kg m$^{-3}$) correspond to 66 ± 7%, 95 ± 15%, and 102 ± 16%, when expressed as a percent of the respective atmospheric densities (0.58, 0.85, and 1.38 kg m$^{-3}$), respectively. The high value at 51.3 km is consistent with the over-estimations predicted by qualitative analysis of the respective calibration curve. As described in **Section 2.1**, $CO_2$ abundances between 51.3 and 45.2 km then decrease by ~$10^4$-fold due to aerosols in middle and lower clouds clogging the LNMS inlets. Across 42.2 to 24.4 km, counts then steadily resume back to near expected values upon clearing of the clogs.

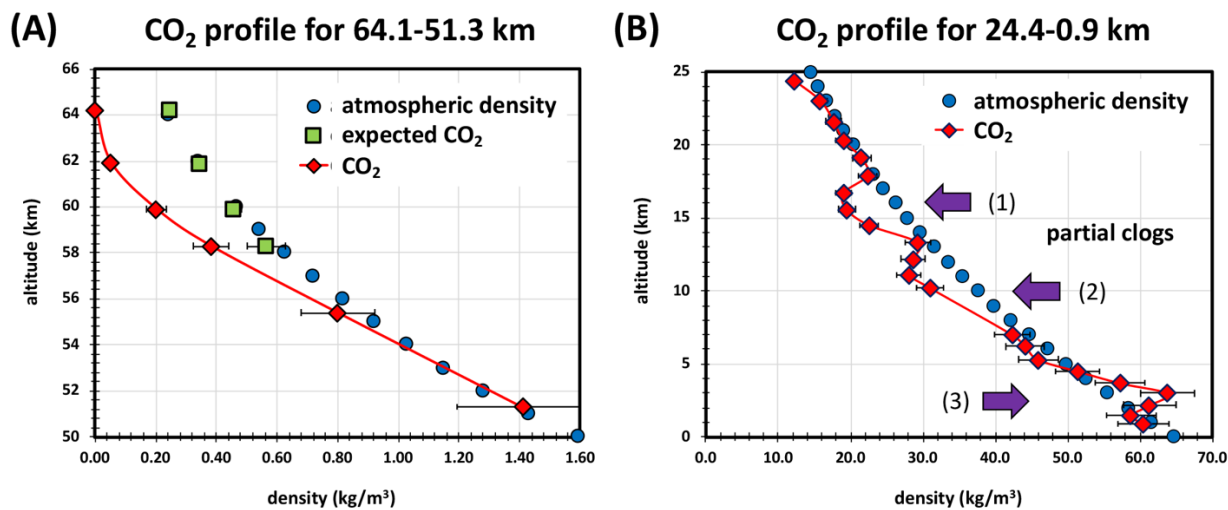

**Fig. 9**. Altitude profiles for $CO_2$ (red diamonds), which were normalized to $^{136}Xe^{2+}$ and expressed in units of density (kg/m³), and atmospheric density (blue circles) for (A) the clouds between 64.2-51.3 km and (B) the lower atmosphere between 24.4-0.9 km, where the green squares represent the expected $CO_2$ densities in between 64.2-58.3 km and block arrows represent (1) the decrease in apparent density at ~15 ± 2 km, (2) the decrease at ~10 ± 3 km, and (3) the increase at ~3 ± 1 km; error bars represent the propagated error.

Between 23.0-17.9 km, after resumption of optimal atmospheric intake, averaged $CO_2$ densities (15.9-22.4 kg m$^{-3}$) are 96.4 ± 5.5% (propagated standard error of the mean; n = 5) of the respective atmospheric values (16.7-23.3 kg m$^{-3}$). Similar relative $CO_2$ abundances are obtained at 13.3 km, 7.0-4.5 km, and 1.5-0.9 km with percent values of 95.2 ± 5.6% (29.3 ± 1.7 kg m$^{-3}$), 96.2 ± 5.5% (46.0 ± 2.0 kg m$^{-3}$), and 96.9 ± 5.7% (56.7 ± 3.6 kg m$^{-3}$), respectively. Across these measures (23.0-17.9, 13.3, 7.0-4.5, and 1.5-0.9 km), which represent altitudes of optimal atmospheric intake, averaged values amount to 96.3 ± 5.7% w/w (propagated standard error of the mean; n = 12). Alternatively, when solely considering variance in the measures, averaged optimal values are 96.3 ± 1.6% w/w (standard deviation; n = 12).

At ≤17 km, as labeled in **Fig. 9B** (purple block arrows), $CO_2$ densities temporarily deviate from the averaged descent value (96.3 ± 1.6%). For example, densities for $CO_2$ decrease by ≤25% at a range defined as ~15 ± 2 km (16.7-14.5 km), decrease by ≤15% at ~10 ± 3 km (12.2-10.2 km), and *increase* up to 18 ± 7% at ~3 ± 1 km (3.7-2.2 km). Similarly, as described in **Section 3.5.2**, the relative $^{136}Xe^{2+}/^{136}Xe^{+}$ ratios show intermittent decreases at altitude ranges (14.5-12.2 km, 10.2-6.2 km, and 3.0-2.2 km) that overlap with the dips in the $CO_2$ density.

We've interpreted these combined observations as the LNMS experiencing partial and rapidly clearing clogs of the primary inlet at ~15 ± 2, ~10 ± 3, and ~3 ± 1 km with optimal $CO_2$ values resuming at 13.3, 7.0-4.5, and 1.5-0.9 km. During the partial clogs, the trapped materials likely vaporized and dissociated in a rapid manner under the high pressures and temperatures to yield an influx of chemical species into the mass analyzer.

We posit that this influx was measured as spikes in the data at differing masses during and after the partial clogs – similar to observations from the main clog. Our review of the data reveals >37 differing mass positions between *m/z* 10-208 that show increases in counts (or spikes) of up to ~500-fold at altitudes corresponding to partial clogs. These combined observations suggest that the data spikes correspond to chemical signatures that represent the elemental and possible molecular composition of a deep atmospheric haze of particles at ≤17 km.

### 3.13 Altitude Profiles for $CO_2$ Density and Particle Extinction Coefficients

We present a comparison of the $CO_2$ density profile to the volumetric scattering extinction coefficients measured by VeNeRa 13 and 14 (Grieger et al., 2004) and the Pioneer Venus Large Probe (Ragent et al., 1985), herein labeled as V13, V14, and PVLP, respectively. As shown in **Fig. 10A**, the largest particle densities or cross sections (extinction coefficients ~3-6 km$^{-1}$) are measured between ~50-46 km in the middle and lower cloud deck, which matches the altitude range (50.3-45.5 km) where the LNMS registers a significant drop in $CO_2$ densities or counts. These observations are consistent with the LNMS experiencing a major clog of the inlets by particles or aerosols from the clouds.

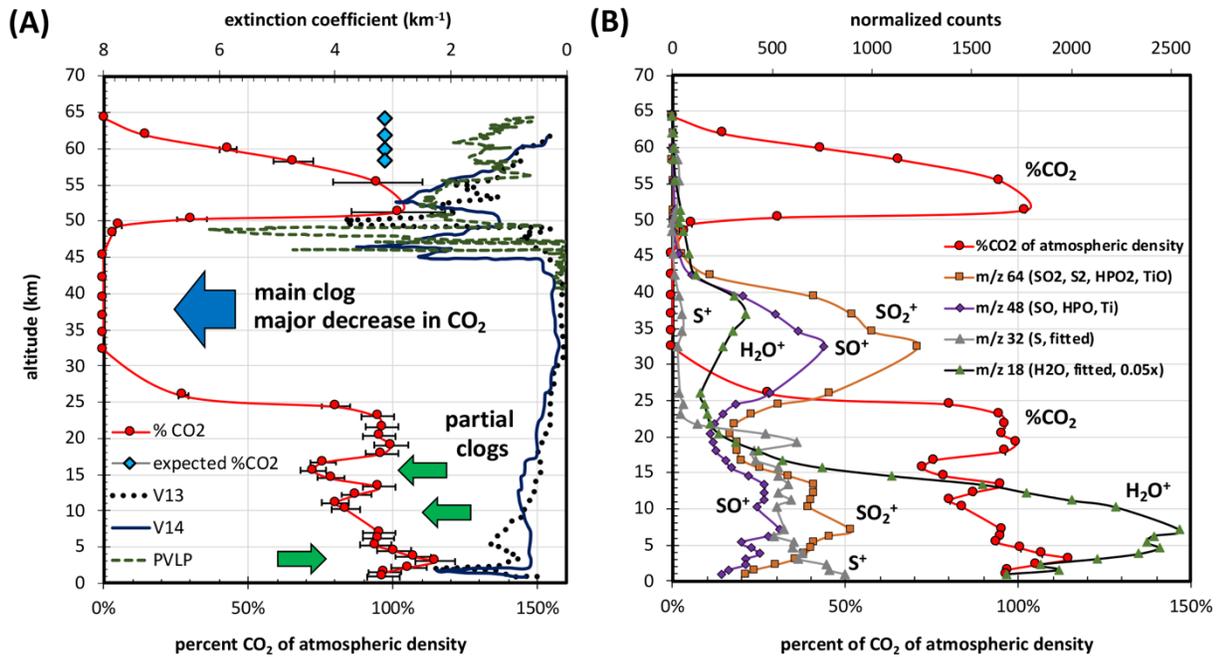

**Fig. 10**. Altitude profiles for %CO$_2$ (relative to atmospheric densities; red circles, red line) plotted against (A) the volumetric scattering coefficients measured from ~62-1 km by VeNeRa 13 (black dotted line) and VeNeRa 14 (thick blue line), and from ~65-40 km by the PV Large Probe (dashed green line), where block arrows represent areas of full (blue) and partial (green) blockage of the inlets, and error bars represent the propagated error; and (B) the altitude profiles for %CO$_2$ and the normalized counts (upper x-axis) at $m/z$ 64 (brown squares; isobars: SO$_2^+$, S$_2^+$, HPO$_2^+$, and/or TiO$^+$), $m/z$ 48 (purple diamonds; isobars: SO$^+$, HPO$^+$, and/or Ti$^+$), $m/z$ 32 (gray triangles; S$^+$, fitted), and $m/z$ 18 (green triangles; H$_2$O$^+$, fitted, plotted at 0.05x for clarity).

Similarly, moderate *increases* in the extinction coefficients at ≤17 km (~0.6-2.0 km$^{-1}$) roughly correspond to where the LNMS experienced the partial clogs at ~15 ± 2, ~10 ± 3, ~3 ± 1 km (**Fig. 10A**). Measures from V14 reveal a sharp increase of ~40% in the extinction coefficients between ~18 km (~0.51 km$^{-1}$) and ~17 km (~0.72 km$^{-1}$), which corresponds to where the partial clogs to the LNMS began. Across ~17-10 km, which overlaps with the partial clogs at ~15 ± 2 and ~10 ± 3 km, the V14 measures maximally increase at ~15 km by ~55% (~0.80 km$^{-1}$) and at ~13 km by ~64% (~0.84 km$^{-1}$), when expressed relative to ~18 km (~0.51 km$^{-1}$). Across ~9-3 km, which overlaps with the partial clog at ~10 ± 3 and ~3 ± 1 km, measures from V13 show a peak increase in the extinction coefficients of ~80% at ~5 km (~1.33 km$^{-1}$), when expressed relative to ~10 km (~0.74 km$^{-1}$). Across ~3-1 km, measures from both V13 and V14 (V13/14) show substantial spikes at ~2 km (~2.24-2.30 km$^{-1}$), which directly overlaps with the partial clog at ~3 ± 1 km, where apparent $CO_2$ densities increase up to ~18 ± 10%.

## 4. Discussion

### 4.1 LNMS Profile for $CO_2$

Our model for the LNMS mass spectral data yields $CO_2$ abundances in units of density (kg m$^{-3}$). Our analysis yields a 10-point detection of $CO_2$ inclusive of vertical fragmentation trends for $CO_2^+$, $CO^{18}O^+$, $^{13}CO_2^+$, $C^{18}O^+$, $^{13}CO^+$, $CO^+$, $^{13}CO_2^{2+}$, $CO_2^{2+}$, $O^+$, and $C^+$. The $^{13}C/^{12}C$ ratio (1.260 x 10$^{-2}$ ± 0.072 x 10$^{-2}$; 55.4-0.9 km, 25 spectra; 5.7% error) obtained from $CO_2$ isotopologues is slightly higher, though within error, to values from the original LNMS investigations (1.19 x 10$^{-2}$ ± 0.06 x 10$^{-2}$; ≤24 km, presumed; 5.0% error), mass spectral studies by VeNeRa 11 and 12 (V11/12; 1.2 x 10$^{-2}$ ± 0.1 x 10$^{-2}$; ≤23 km; 8.3% error), and remote analyses of $CO_2$ isotopologues (1.2 x 10$^{-2}$ ± 0.2 x 10$^{-2}$; >60 km, approximately; 17% error) (Hoffman, Hodges, Donahue, et al., 1980, Istomin et al., 1979, Johnson and de Oliveira, 2019).

The $^{18}O/^{16}O$ ratio (2.195 x 10$^{-3}$ ± 0.156 x 10$^{-3}$; 55.4-0.9 km, 25 spectra; 7.1% error) obtained from $CO_2$ isotopologues is higher, though within error, to values from the original LNMS investigations (2.0 x 10$^{-3}$ ± 0.1 x 10$^{-3}$; ≤24 km, presumed; 5.0% error) and remote analyses of $CO_2$ (2.0 x 10$^{-3}$ ± 0.3 x 10$^{-3}$; >60 km, approximately; 15% error). To our knowledge, V11/12 did not measure the $^{18}O/^{16}O$ ratio (Hoffman, Hodges, Donahue, et al., 1980, Istomin et al., 1979, Johnson and de Oliveira, 2019).

The altitude profile for $CO_2$ reveals a steep, non-linear gradient, where densities increase by ~76-fold across the clouds (55.4 km, 0.80 kg m$^{-3}$) towards the surface (0.9 km, 60.46 kg m$^{-3}$). The vertical profile for $CO_2$ is consistent with hydrostatic equilibrium to yield a scale height of 16.7 ± 0.9 km ($R^2$ = 0.998) between 23.0-0.9 km (assuming uncertainties of ~5.5% in the densities). We obtain a similar scale height of 17.2 ± 0.2 km ($R^2$ = 0.999) from Venus atmospheric densities between 24-1 km, which were calculated from pressures and temperatures, and assume a ≤1.2% error at <40 km, per Seiff et al. (1985). The LNMS scale height for $CO_2$ (23.0-0.9 km) also follows the trend of larger $CO_2$ scale heights towards the surface in Hunten (1968) and falls within the broad range of 0-15 km reported for Venus scale heights in Bézard et al. (2009).

The current altitude profile for $CO_2$ at Venus, when expressed in units of density (kg m$^{-3}$) or volumetric percent (% v/v), include measures across ~160-80 km (~$10^{-11}$-$10^{-4}$ kg m$^{-3}$) obtained through stellar occultation (Piccialli et al., 2015, Vandaele et al., 2016), volumetric percent values obtained by the PV Gas Chromatograph at 51.6 km (95.4 ± 2.5 % v/v), 41.7 km (95.9 ± 0.7 % v/v), and 21.6 km (96.4 ± 0.1 % v/v) (Oyama et al., 1980), volumetric percent values obtained by V11/12 at ≤23 km (95.5% v/v) (Istomin et al., 1979, Istomin et al., 1980), and the volumetric percent values obtained by VeNeRa 4 (V4) at ≤23 km (90-95 % v/v) (Reese and Swan, 1968). In comparison, we've extracted 14 measures of $CO_2$ density between 55.4-0.9. Hence, our results meaningfully update the current altitude profile for $CO_2$.

When excluding the main and partial clogs, we obtain $CO_2$ abundances of 96.6 ± 8.0% w/w between 55.4-0.9 km (propagated standard error of the mean; n = 14); 96.3 ± 5.7% w/w between 23.0-0.9 km (propagated standard error of the mean; n = 12); and 96.9 ± 5.7% w/w between 1.5-0.9 km (propagated standard error of the mean; n = 2). Standard errors for these measures, which represent uncertainties from the calibration curves (*e.g.*, conversion of amps to density; **Fig. 7**), could theoretically decrease upon future access to $CO_2$ control experiments in restricted LNMS documentation. When considering standard deviations across the altitude profile, averaged values show limited variances of 96.6 ± 2.4% w/w across 55.4-0.9 km (standard deviation; n = 14); 96.3 ± 1.6% w/w across 23.0-0.9 km (standard deviation; n = 12); and 96.9 ± 0.4% w/w across 1.5-0.9 km (standard deviation; n = 2).

During descent of the PVLP, $CO_2$ abundances were measured by the LNMS and PVGC. Our volumetric percent value for $CO_2$ from the LNMS data at 21.6 km (96.3 ± 5.7% w/w, propagated standard error) is comparable in both magnitude and error to the PVGC value at 21.6 km ($97.7^{+4.1}_{-5.9}$ % w/w) – which considers errors in Oyama et al. (1980) that are listed with a 3 σ confidence interval. The LNMS-derived abundances across 55.4-0.9 km of 96.6 ± 2.4% w/w (standard deviation; n = 14) – when excluding the main and partial clogs – are also within error to the bulk averaged value of 97.4 ± 0.3% w/w (standard deviation) for the total measurements from the PVGC (at 51.6, 41.7, and 21.6 km).

When considered together, these comparisons validate our model and interpretation of the LNMS data. Our $CO_2$ volumetric abundances match prior measures obtained *in situ*, the $CO_2$-derived $^{13}C/^{12}C$ and $^{18}O/^{16}O$ ratios match prior measures and are expressed with higher significant figures, and $CO_2$ fragmentation yields are similar to terrestrial reference values (NIST and published cross sections). Accordingly, through this model, we extracted a 14-point altitude profile for $CO_2$ in units of density across 55.4-0.9 km, which is perhaps the most detailed altitude profile for $CO_2$ among *in situ* measures, to date, and the first $CO_2$ profile for the lower atmosphere (among *in situ* measures) to be expressed as density.

### *4.2 Performance of the LNMS During Descent*

We tracked several internal standards in the data ($H^+$, $C^+$, $CH_2^+$, $CH_3^+$, $CO^+$, $N_2^+$, $^{40}Ar^+$, $^{136}Xe^{2+}$, and $^{136}Xe^+$) to obtain insights into the impacts of descent and mass on the ionization yields, fragmentation yields, and peak shapes. The combined diagnostic trends support nominal operations for the mass analyzer throughout the descent across the clouds (58.8-51.3 km), during the main clog (~50-25 km), and across the high pressure and temperature environment of the lower atmosphere (24.4-0.9 km).

Pre-sampling spectra indicate repeatable electron ionization yields (*e.g.*, $^{136}Xe^{n+}$ and $^{40}Ar^{n+}$) and repeatable fragmentation yields (*e.g.*, ionization of $CH_4$ to yield $CH_4^+$, $CH_3^+$, $CH_2^+$, $C^+$, and $H^+$). Descent spectra (*i.e.*, during sampling) indicate reproducible electron ionization yields and fragmentation yields when considering the impacts of mass flow intake rates (*e.g.*, before and after Valve 1 closure) on transmission efficiencies, which slightly increase with decreasing *m/z* along the descent.

Mogul et al., *Icarus **Accepted*** (Nov. 2022)

In the descent spectra, the altitude profiles for the FWHM, mass error, and fitted counts of the internal standards (peak fitting regression terms) reveal no evidence of systematic noise and no major influences from atmospheric pressure or temperature. Instead, the descent trends indicate reasonable mass accuracies and reproducible peak shapes across the descent profile.

Performance of the LNMS was additionally assessed by tracking the $CO_2$ profile across the descent (densities or counts). As detailed, the $CO_2$ profile reveals evidence of a significant clog at ~50 km (main clog), which corresponds to where the PVLP and V13/14 spacecraft measured substantial increases in particle densities (or cross sections). These observations, along with the ensuing data spikes for $SO_2^+$, $SO^+$, $S^+$, and $H_2O^+$, are consistent with the interpretation that the LNMS experienced a clog of the inlet due to particles (hydrated sulfuric acid) from the middle and lower clouds, which subsequently degraded at the inlet to yield an influx of mass signals in the LNMS.

Analogously, the $CO_2$ profile reveals evidence of several partial and rapidly clearing clogs in the lower atmosphere (≤17 km) at the altitude ranges of ~15 ± 2, ~10 ± 3, and ~3 ± 1 km. The partial clogs roughly match where the V13/14 spacecraft measured moderate increases in particle densities (or cross sections). The partial clogs additionally correspond to the altitudes where several spikes in the LNMS data are observed – similar to trends observed after the main clog.

Accordingly, we've interpretated the data spikes at ≤17 km as mass signatures, in part, of haze particles that partly clogged and degraded at the LNMS inlet to yield an influx of chemical species into the LNMS mass analyzer. Our analyses show that >37 mass positions between *m/z* 10-208 (out of a total 232 mass positions) exhibit spikes in the counts of ~5-500-fold. However, the internal standards show no evidence of data spikes at ≤17 km and, as described, exhibit nominal behavior across the descent at the mass positions (and/or peaks) of *m/z* 1, 12, 14, 15, 28, 40, 68, and 136. Therefore, trends from the internal standards negate random noise and/or failures in the mass analyzer as underlying reasons for the data spikes.

We additionally note that data from the lower atmosphere (≤24 km) was explicitly used in the original LNMS investigations. In Hoffman, Hodges, Donahue, et al. (1980), the mixing ratios of $^{36}$Ar, $H_2O$, $H_2S$, and $SO_2$ were determined using the data from <28 km. In Donahue et al. (1981), the mixing ratios of Xe and Kr were obtained at ≤20 km. In Hoffman et al. JGR 1980, the spikes in $SO_2^+$ at ≤6 km (as observed in **Fig. 1**) were very briefly described and noted to be similar to results obtained from ground-based studies, which show formation of $SO_2^+$ and $SO^+$ after dissociation of sulfuric acid at the LNMS inlet under equivalent conditions – which is consistent with our assessment that the data spikes at ≤17 km arise from degraded deep atmospheric haze particles.

In our review of the primary literature over the last 40+ years, we find no criticisms that question the validity or veracity of the published LNMS-derived mixing ratios and, by extension, the LNMS dataset. Hence, our tracking the internal standards, along with the original LNMS investigations, indicate that the data at ≤24 km is of sufficient high quality for in-depth future analyses.

*4.3 Methane in the LNMS Data*

To explain the unexpected and substantial increases in methane ($CH_4$) across the descent (64.2-0.2 km), we tracked all species related to $CH_4^+$, compared the trends to the profiles for $CO_2^+$, $CO^+$, $O^+$, and $C^+$, constructed a preliminary conversion of counts to mixing ratios for $CH_4$, and considered the behavior of the LNMS during operation. Our assessments of nominal LNMS operations negate failures in the mass analyzer as underlying reasons for the large increases in $CH_4^+$ (and related products). Rather, our review of the literature and LNMS data suggest that the large increases in counts are due to an influx of methane formed external to the LNMS due to spacecraft degradation and appreciable behavior of methane as a non-gettered species within the LNMS mass analyzer (*e.g.*, appreciable rates of desorption from the getter).

As described In Fimmel et al. (1995), the suite of PV probes experienced common sensor and electrical anomalies at ≤12.5 km. Explanations included breakdown of the Kapton film (an aromatic polyimide) that wrapped the external harnesses or sets of wires for external components on the spacecraft such as the atmospheric temperature sensors. As described in Seiff et al. (1995), exposures to sulfuric acid significantly degrade Kapton films under high pressure and temperature (60 min exposures to 96% v/v $H_2SO_4$ at 100 bar and 760 K in $N_2$).

In Seiff et al. (1995) it is further posited that the aromatic C-H groups of the Kapton film would oxidize in the presence of atmospheric $CO_2$ under the conditions at ≤12.5 km (≥40 bar, ≥630 K). We note that concomitant reduction of $CO_2$ by the aromatic C-H groups from the Kapton insulation would lead to the formation of $CH_4$. We suggest that the damaged external harnesses that potentially liberated $CH_4$ were proximal to the LNMS inlets and that imbibing of the $CO_2$-derived $CH_4$ resulted in the observed trends from the lower atmosphere (≤24 km).

When considering mixing ratios of external $CH_4$ across the descent, our preliminary model yields, ~15 ± 12 ppm at 51.3 km, ~130 ± 100 ppm at 21.6 km, and ~1200 ± 1000 ppm at 0.9 km; these values were calculated against $CO_2$ and $N_2$ abundances, corrected for the apparent *m/z*-dependent impacts on getter efficiency, and expressed with an error representing the upper and lower outcomes of the corrections. In the original LNMS investigations (Donahue and Hodges, 1993), the potential $CH_4$ mixing ratios were calculated against $^{36}Ar$ and reported as ~980 ppm at 51.3 km and ~2800 ppm at 0.9 km, which are respectively ~65 and ~2-fold higher than our values. This is significant since the PVGC (Oyama et al., 1980) and gas chromatographs on V13/14 (Moroz, 1983) did not detect $CH_4$ in Venus' atmosphere, which effectively placed upper limits of ~10 ppm at 51.6 km (PVGC), ~0.6 ppm at 21.6 km (PVGC), and ~0.5 ppm across 58-3.5 km (presumed range; VeNeRa).

In comparison, our $CH_4$ mixing ratio in clouds at 51.3 km (~15 ± 10 ppm) is within error to the PVGC upper limits (10 ppm), which validates our preliminary estimates of getter efficiency and conversions of counts to mixing ratios. However, for the lower atmosphere at 21.6 km, our mixing ratio is ~220-fold *higher* than the PVGC upper limit, which perhaps represents overestimation of the apparent getter efficiency for $CH_4$ as mass flow intake rates increased across the lower atmosphere. Alternatively, our $CH_4$ high mixing ratios at 21.6 km may be due to the damaged external harnesses (where the $CH_4$ was formed) being proximal to the LNMS inlets and further away from the PVGC inlet. Current efforts are aimed at ascertaining the locations of the external harnesses on the PVLP.

We also considered several other possible sources of $CH_4$. Reduction of atmospheric CO is excluded as the major source of $CH_4$ since the mixing ratios of ~30 and 20 ppm at 51.6 and 21.6 km (Oyama et al., 1980) provide insufficient mass balance. Hydrogenation of $CO^+$ and/or $C^+$ (arising from $CO_2^+$) within the LNMS is likely not a major source of $CH_4^+$ since our analyses cast doubt on the presence of substantial $H_2$ (*m/z* 2) in the spectrometer walls, as was postulated in Donahue et al. (1997). Instead, our analyses are supportive – *though not definitive* – of counts at *m/z* 2 arising from $^4He^{2+}$.

We also find no viable carbon sources – within the LNMS – to account for internal hydrogenation to yield $CH_4^+$ in high counts (~$2.7 \times 10^7$ total counts; fitted). For example, the descent profiles for $CO^+$ and $C^+$ arising from $CO_2$ ionization and fragmentation (when expressed relative to $CO_2^+$) show no losses to commensurately support the *massive* increases in $CH_4^+$. Also, the extremely high internal counts of $CH_4^+$ potentially negate the spectrometer walls as major sources of carbon; nonetheless, the available LNMS literature do not indicate if carbon-containing steel (*e.g.*, stainless steel) was used in the construction of the mass analyzer.

Other unlikely carbon sources for $CH_4$ – within the LNMS – include sealants (*e.g.*, xylene) and residues from bake out procedures. We estimate that a *minimum* of ~0.11 g of atomic carbon (~0.21 g/L), or ~0.12 g of xylene (~0.23 g/L), are required to yield the total $CH_4^+$ observed across the descent (~$2.7 \times 10^7$ total counts). We obtained these mass estimates by (1) summing counts for $CH_4^+$ across the entire descent (~$2.7 \times 10^7$ counts, 64.2-0.2 km, 39 spectra), (2) correcting for $CH_4^+$ arising from the internal calibrant gas across all spectra in the descent profile using averaged counts from the pre-sampling data, (3) converting total corrected $CH_4^+$ counts (~$2.6 \times 10^7$ counts)

Mogul et al., *Icarus* **Accepted** (Nov. 2022)

to internal mixing ratios (~4.2x10$^5$ internal ppm) with respect to total $CO_2^+$ and $CH_4^+$ (the highest abundant species in the data), (4) conversion of the $CH_4^+$ internal mixing ratio (~4.2x10$^5$ internal ppm) to density using the equation of state, (5) assuming an internal mass analyzer volume that is 5% of the total LNMS volume (10.6 L), and (6) assuming quantitative carbon recovery across (A) atomization of the carbon source, (B) adsorption onto the LNMS walls, (C) desorption from the walls, (D) hydrogenation to yield $CH_4$, (E) ionization to yield $CH_4^+$, and (F) detection at the electron multiplier. Under these assumptions, such levels of organic contamination for a spacecraft mass spectrometer are likely improbable.

When considering $CH_4$ in the clouds (64.2-51.3 km), the reduction of $CO_2$ (or CO) by the Kapton insulation is unlikely given the mild pressures and temperatures between 64.2-51.3 km (~0.1-0.9 bar, ~235-335 K). Nevertheless, the external temperature sensor (T1) on the PVLP, as shown in Figure 3(a) in Seiff et al. (1995), temporarily failed (or shorted) between ~58-37 km (or between the temperatures of ~280-440 K). Further, our review of the LNMS data shows that counts for $CH_4^+$, upon sampling at 64.2 km, moderately increase to yield a potential ~15 ± 10 ppm external $CH_4$ at 51.3 km. We venture that the external harnesses may have initially degraded in the clouds (64.2-51.3 km) after reactions with acid vapors such as HCl (<1 ppm) and $H_2SO_4$ (<1 ppm) (Johnson and de Oliveira, 2019, Oschlisniok et al., 2021) and/or the particles possessing appreciable extinction coefficients (~1.5-3 km$^{-1}$; **Fig. 10A**). However, in the absence of suitable spacecraft-associated hydride sources under the mild conditions of the clouds, alternative sources for $CH_4^+$ include undocumented cloud-based chemistry, where thermodynamic models suggest a maximum of <0.1 ppm $CH_4$ at 30 km (Pollack et al., 1993).

## 5. Conclusions

We extracted novel information pertaining to Venus' atmosphere by thorough re-analysis of archived mass spectral data acquired by the Pioneer Venus Large Probe Neutral Spectrometer (LNMS). In this study, we developed an analytical model for the LNMS data that accounts for the impacts of descent and changes instrument configuration and enables the disentanglement of isobaric species via a statistical data fitting routine that adjusts for mass-dependent changes in peak widths and centers at each altitude. To ultimately yield tangible mixing ratios and altitude profiles for atmospheric gases, we focused our study on extracting the $CO_2$ altitude profile, converting the $CO_2$ counts to units of density, and characterizing instrumental output during descent. In total, our analyses yield 14 measures of $CO_2$ density between 55.4 km (middle clouds) and 0.9 km (96.6 ± 2.4% w/w, standard deviation; 96.6 ± 8.0% w/w, propagated error), which represents the most complete altitude profile for $CO_2$ at ≤60 km to date.

Tracking of the $CO_2$ density across the descent also reveals hitherto unreported partial and rapidly clearing clogs of the LNMS inlet in the lower atmosphere at ~15 ± 2, ~10 ± 3, and ~3 ± 1 km (or ≤17 km). Our review of the LNMS data indicate that the partial clogs were followed by several spikes (or temporary increases) in the counts across several mass positions between *m/z* 1-208. We've interpreted these combined observations as evidence of a deep atmospheric haze of particles at ≤17 km, which possess sufficient densities (or sizes) to impact atmospheric intake, and sufficient labilities to degrade at the LNMS inlet to yield the observed influx of mass signals.

The presence of a deep atmospheric haze is consistent with several reports in the Venus literature. For example, measurements from the PV night and north probes, which showed increases in the backscattering cross sections (Ragent 1980), were interpreted in Seiff et al. (1995) as support for a particle-bearing layer between ~5.4-7.1 km, which in turn was likely referred to as a "gloomy red murk" in Figure 5.-12 in Fimmel et al. (1995). Similarly, in Grieger et al. (2004), cross sections obtained from radiance measurements from V13/14 at <5 km were interpreted as evidence of a "near surface cloud layer". In Anderson (1969), measures from Mariner V and V4 were interpreted as support for dust particles at ≤25 km. We add that the PV LNMS may have obtained evidence of particles – or a deep atmospheric haze – at ≤17 km.

In conclusion, this study yields a robust analytical model for future interrogations of the LNMS dataset and highlights the remarkable design of the LNMS. Future efforts are focused on extracting the altitude profile for $N_2$, the cloud profiles for $H_2O$, altitude profiles for neutral gases, including the noble gases, expression of mixing ratios against $CO_2$ and $N_2$ (as opposed to those derived from $^{36}Ar$ from the original investigations), and elucidation of the chemical composition of the deep atmospheric haze. Lastly, our work points to potential operational hurdles for future atmospheric gas sampling studies at <20 km, such as those proposed by the upcoming DAVINCI mission (Garvin et al., 2022), which could be impacted by particles of the deep atmospheric haze.

**Data Availability**

The LNMS archive data are available online at the NASA Space Science Data Coordinated Archive (NSSDC) (https://nssdc.gsfc.nasa.gov/nmc/dataset/display.action?id=PSPA-00649).

**Acknowledgements**

We acknowledge Paul Rimmer for helpful discussions. RM acknowledges administrative support from the Blue Marble Institute of Science and funding from the National Aeronautics and Space Administration (NASA) Nexus for Exoplanet System Science (NExSS) award (80NSSC21K1176). SSL was supported by NASA (NNX16AC79G). MJW was supported by the NASA Astrobiology Program through collaborations arising from his participation in the NExSS and the NASA Habitable Worlds Program. MJW also acknowledges support from the GSFC Sellers Exoplanet Environments Collaboration (SEEC), which is funded by the NASA Planetary Science Division's Internal Scientist Funding Model.

**References**

Adamczyk, B., Boerboom, A., Schram, B., Kistemaker, J., 1966. Partial ionization cross sections of He, Ne, $H_2$, and $CH_4$ for electrons from 20 to 500 eV. J. Chem. Phys. 44**,** 4640-4642.

Anderson, A. D., 1969. Dust in the lower atmosphere of Venus. Science. 163**,** 275-276.

Antonita, T. M., Kumar, K. P., Das, T. P., 2022. Overview of ISRO's future Venus Orbiter Mission. 44th COSPAR Scientific Assembly. Held 16-24 July. 44**,** 342.

Belov, V., Ustinov, Y. K., Komar, A., 1978. Carbon monoxide and carbon dioxide interaction with tantalum. Surface Sci. 72**,** 390-404.

Bézard, B., Tsang, C. C., Carlson, R. W., Piccioni, G., Marcq, E., Drossart, P., 2009. Water vapor abundance near the surface of Venus from Venus Express/VIRTIS observations. J. Geophys. Res. Planets. 114.

Bourcey, N., et al., 2008. Leak-tight welding experience from the industrial assembly of the LHC cryostats at CERN. AIP Conference Proceedings, Vol. 985. American Institute of Physics, pp. 325-332.

Brenton, A. G., Godfrey, A. R., 2010. Accurate Mass Measurement: Terminology and Treatment of Data. J. Am. Soc. Mass Spectrom. 21**,** 1821-1835.

Carlson, T. A., Nestor Jr, C., Wasserman, N., McDowell, J., 1970. Calculated ionization potentials for multiply charged ions. Atomic Data Nucl. Data Tables. 2**,** 63-99.

Denifl, S., et al., 2002. Multiple ionization of helium and krypton by electron impact close to threshold: appearance energies and Wannier exponents. J. Phys. B. 35**,** 4685.

Dibeler, V., Mohler, F. L., 1950. Mass spectra of the deuteromethanes. J. Research Nat. Bur. Standards. 45**,** 441-444.

Doak, G., Gilbert Long, G., Freedman, L. D., 2000. Arsenic compounds. Kirk-Othmer Encyclopedia of Chemical Technology.

Dolder, K., Harrison, M., Thonemann, P., 1961. A measurement of the ionization cross-section of helium ions by electron impact. Proc. Royal Soc. London A. 264**,** 367-378.

Donahue, T., Hoffman, J., Hodges Jr, R., 1981. Krypton and xenon in the atmosphere of Venus. Geophys. Res. Letters. 8**,** 513-516.

Donahue, T. M., Hoffman, J. H., Hodges, R. R., Watson, A. J., 1982. Venus was wet - A measurement of the ratio of deuterium to hydrogen. Science. 216**,** 630-633.

Donahue, T. M., Hodges, R. R., Jr., 1992. Past and present water budget of Venus. J. Geophys. Res. 97**,** 6083-6091.

Donahue, T. M., Hodges, R. R., 1993. Venus methane and water. Geophys. Res. Letters. 20, 591-594.

Donahue, T. M., Grinspoon, D. H., Hartle, R. E., Hodges Jr, R. R., 1997 Ion/neutral escape of hydrogen and deuterium: Evolution of water. In: D. M. Hunten, R. J. Phillips, S. W. Bougher, (Eds.), Venus II--geology, Geophysics, Atmosphere, and Solar Wind Environment. University of Arizona Press, pp. 385-414.

Fimmel, R. O., Colin, L., Burgess, E., 1995. Pioneering Venus: A Planet Unveiled. NASA.

Fox, R., 1959 Study of Multiple Ionization in Helium and Xenon by Electron Impact. Advances in Mass Spectrometry. Elsevier, pp. 397-412.

Fox, R., 1960. Multiple ionization in argon and krypton by electron impact. J. Chem. Phys. 33**,** 200-205.

Garvin, J. B., et al., 2022. Revealing the Mysteries of Venus: The DAVINCI Mission. Planet. Sci. 3**,** 117.

Ghail, R., 2021. How will EnVision improve our understanding of Venus? , AAS/Division for Planetary Sciences Meeting Vol. 53, pp. 315.02.

Gluch, K., et al., 2003. Cross sections and ion kinetic energies for electron impact ionization of $CH_4$. Int. J. Mass Spectrom. 228**,** 307-320.

Grieger, B., Ignatiev, N., Hoekzema, N., Keller, H. U., 2004. Indication of a near surface cloud layer on Venus from reanalysis of Venera 13/14 spectrophotometer data. Proc. Int. Workskhop "Planetary Probe Atmospheric Entry and Descent Trajectory Analysis and Science". 544**,** 63-70.

Habibzadeh, F., Habibzadeh, P., 2015. How much precision in reporting statistics is enough? Croat. Med. J. 56**,** 490.

Harrison, D., Barry, T., Turner, G., 2004. Possible diffusive fractionation of helium isotopes in olivine and clinopyroxene phenocrysts. Euro. J. Mineral. 16**,** 213-220.

Haynes, W. M., 2016. CRC Handbook of Chemistry and Physics. CRC Press, Boca Raton, Florida.

Hoffman, J. H., Hodges, R. R., Duerksen, K. D., 1979. Pioneer Venus large probe neutral mass spectrometer. J. Vac. Sci. Technol. 16**,** 692-694.

Hoffman, J. H., Hodges, R. R., McElroy, M. B., Donahue, T. M., Kolpin, M., 1979. Composition and structure of the Venus atmosphere: Results from Pioneer Venus. Science. 205**,** 49-52.

Hoffman, J. H., Hodges, R. R., McElroy, M. B., Donahue, T. M., Kolpin, M., 1979. Venus lower atmospheric composition: Preliminary results from Pioneer Venus. Science. 203**,** 800-802.

Hoffman, J. H., Hodges, R. R., Donahue, T. M., McElroy, M. B., 1980. Composition of the Venus lower atmosphere from the Pioneer Venus mass spectrometer. J. Geophys. Res. Space Phys. 85**,** 7882-7890.

Hoffman, J. H., Hodges, R. R., Wright, W. W., Blevins, V. A., Duerksen, K. D., Brooks, L. D., 1980. Pioneer Venus sounder probe neutral gas mass spectrometer. IEEE Trans. Geosci. Remote Sens. GE-18**,** 80-84.

Hoffman, J. H., Oyama, V. I., Von Zahn, U., 1980. Measurements of the Venus lower atmosphere composition: A comparison of results. J. Geophys. Res. Space Phys. 85**,** 7871-7881.

Hunten, D. M., 1968. The structure of the lower atmosphere of Venus. J. Geophys. Res. 73**,** 1093-1095.

Inoue, M., Watanabe, T., Danno, A., 1964. Isotope Effect on Fragmentation and Kinetic Energy of Fragment Ions from Carbon Dioxide Produced by Electron Impact. Jap. J. Appl. Phys. 3**,** 761.

Istomin, V., Grechnev, K., Kochnev, V., Ozerov, L., 1979. Composition of the Venus low atmosphere according to the data of mass spectrometers. Kosmicheskie Issledovaniya. 17**,** 703-707.

Istomin, V., Grechnev, K., Kotchnev, V., 1980. Mass spectrometer measurements of the composition of the lower atmosphere of Venus. COSPAR Colloquia Series, Vol. 20. Elsevier, pp. 215-218.

Johnson, N. M., de Oliveira, M. R., 2019. Venus atmospheric composition in situ data: a compilation. Earth Space Sci. 6**,** 1299-1318.

Kahlert, H., Huber, B., Wieseman, K., 1983. Charge exchange and transfer ionisation in low-energy Ne2+-Xe collisions. J. Phys. B. 16**,** 449.

King, S. J., Price, S. D., 2008. Electron ionization of $CO_2$. Int. J. Mass Spectrom. 272**,** 154-164.

Kramida, A., Ralchenko, Y., Reader, J., (2021), N. A. T., 2022. NIST Atomic Spectra Database (version 5.9). National Institute of Standards and Technology Gaithersburg, MD.

Lewis, J. S., Fegley, B., 1982. Venus: Halide cloud condensation and volatile element inventories. Science. 216**,** 1223-1225.

Linkin, V., et al., 1986. Vertical thermal structure in the Venus atmosphere from Provisional VEGA-2 temperature and pressure data. Sov. Astron. Let. 12**,** 40-42.

Liu, X., Shemansky, D. E., 2006. Analysis of electron impact ionization properties of methane. J. Geophys. Res. Space Phys. 111.

Mishima, K., et al., 2018. Accurate determination of the absolute $^3$He/$^4$He ratio of a synthesized helium standard gas (helium standard of Japan, HESJ): toward revision of the atmospheric $^3$He/$^4$He ratio. Geochem. Geophys. Geosyst. 19**,** 3995-4005.

Mogul, R., Limaye, S. S., Way, M., Cordova, J. A., 2021. Venus' mass spectra show signs of disequilibria in the middle clouds. Geophys. Res. Letters. 48**,** e2020GL091327.

Moroz, V., 1983 Summary of preliminary results of the Venera 13 and Venera 14 missions. In: D. M. Hunten, (Ed.), Venus. University of Arizona Press, Tucson, AZ, pp. 45-68.

Nic, M., Jirat, J., Kosata, B., 2006. Compendium of Chemical Terminology International Union of Pure and Applied Chemistry, https://goldbook.iupac.org/.

Oschlisniok, J., et al., 2021. Sulfuric acid vapor and sulfur dioxide in the atmosphere of Venus as observed by the Venus Express radio science experiment VeRa. Icarus. 362**,** 114405.

Oyama, V. I., Carle, G. C., Woeller, F., Pollack, J. B., Reynolds, R. T., Craig, R. A., 1980. Pioneer Venus gas chromatography of the lower atmosphere of Venus. J. Geophys. Res. 85**,** 7891-7902.

Ozima, M., Takayanagi, M., Zashu, S., Amari, S., 1984. High $^3$He/$^4$He ratio in ocean sediments. Nature. 311**,** 448-450.

Piccialli, A., et al., 2015. Thermal structure of Venus nightside upper atmosphere measured by stellar occultations with SPICAV/Venus Express. Planet. Space Sci. 113**,** 321-335.

Pollack, J. B., et al., 1993. Near-infrared light from Venus' nightside: A spectroscopic analysis. Icarus. 103**,** 1-42.

Prohaska, T., et al., 2022. Standard atomic weights of the elements 2021 (IUPAC Technical Report). Pure Appl. Chem. online May 4, 2022.

Ragent, B., Esposito, L. W., Tomasko, M. G., Marov, M. Y., Shari, V. P., Lebedev, V. N., 1985. Particulate matter in the Venus atmosphere. Adv. Space Res. 5**,** 85-115.

Reese, D. E., Swan, P. R., 1968. Venera 4 probes atmosphere of Venus. Science. 159**,** 1228-1230.

Rejoub, R., Lindsay, B., Stebbings, R., 2002. Determination of the absolute partial and total cross sections for electron-impact ionization of the rare gases. Phys. Review A. 65**,** 042713.

Schram, B., Boerboom, A., Kleine, W., Kistemaker, J., 1966. Amplification factors of a particle multiplier for multiply charged noble gas ions. Physica. 32**,** 749-761.

Scott, N., Carter, D. E., Fernando, Q., 1989. Reaction of gallium arsenide with concentrated acids: formation of arsine. Am. Industrial Hyg. Assoc. J. 50**,** 379-381.

Seiff, A., et al., 1980. Measurements of thermal structure and thermal contrasts in the atmosphere of Venus and related dynamical observations: Results from the four Pioneer Venus probes. J. Geophys. Res. Space Phys. 85**,** 7903-7933.

Seiff, A., et al., 1985. Models of the structure of the atmosphere of Venus from the surface to 100 kilometers altitude. Adv. Space Res. 5**,** 3-58.

Seiff, A., et al., 1995. Pioneer Venus 12.5 km Anomaly Workshop Report (Volume I). NASA, Ames Research Center.

Sigaud, L., Montenegro, E. C., 2015. Cross sections for the $CH_2$ formation pathways via $CH_4$ fragmentation by electron impact. J. Phys. B. 48**,** 115207.

Smith, D., Adams, N., Alge, E., Villinger, H., Lindinger, W., 1980. Reactions of $Ne^{2+}$, $Ar^{2+}$, $Kr^{2+}$ and $Xe^{2+}$ with the rare gases at low energies. J. Phys. B. 13**,** 2787.

Smrekar, S., et al., 2022. VERITAS (Venus emissivity, radio science, InSAR, topography, and spectroscopy): a discovery mission. 2022 IEEE Aerospace Conference (AERO). IEEE, pp. 1-20.

Stephan, K., Märk, T., 1984. Absolute partial electron impact ionization cross sections of Xe from threshold up to 180 eV. J. Chem. Phys. 81**,** 3116-3117.

Straub, H., Renault, P., Lindsay, B., Smith, K., Stebbings, R., 1995. Absolute partial and total cross sections for electron-impact ionization of argon from threshold to 1000 eV. Phys. Review A. 52**,** 1115.

Straub, H., Lindsay, B., Smith, K., Stebbings, R., 1996. Absolute partial cross sections for electron-impact ionization of $CO_2$ from threshold to 1000 eV. J. Chem. Phys. 105**,** 4015-4022.

Sundararajan, V., 2021 Tradespace Exploration of Space System Architecture and Design for India's Shukrayaan-1, Venus Orbiter Mission. ASCEND 2021, pp. 4103.

Takayanagi, M., Ozima, M., 1987. Temporal variation of $^3He/^4He$ ratio recorded in deep-sea sediment cores. J. Geophys. Res. Solid Earth. 92**,** 12531-12538.

Vandaele, A. C., et al., 2016. Contribution from SOIR/VEX to the updated Venus International Reference atmosphere (VIRA). Adv. Space Res. 57**,** 443-458.

Ward, M. D., King, S. J., Price, S. D., 2011. Electron ionization of methane: The dissociation of the methane monocation and dication. J. Chem. Phys. 134**,** 024308.

Weisstein, E. W., 2002. Gaussian function. https://mathworld.wolfram.com/GaussianFunction.html.